\documentclass[conference,10pt]{IEEEtran}

\usepackage{amsmath,amsfonts,amssymb}
\usepackage[ruled,linesnumbered]{algorithm2e}
\usepackage{array}
\usepackage{textcomp}
\usepackage{stfloats}
\usepackage{url}
\usepackage{verbatim}
\usepackage{graphicx}
\usepackage{cite}
\usepackage{xcolor}

\usepackage[colorlinks=true, allcolors=blue]{hyperref}
\usepackage{booktabs}
\usepackage{cleveref}
\usepackage{caption}
\usepackage{subcaption}
\usepackage{footmisc} 

\title{NCorr-FP: A Neighbourhood-based Correlation-preserving Fingerprinting Scheme for Intellectual Property Protection of Structured Data}

\author{
    \IEEEauthorblockN{Tanja Šarčević}
    \IEEEauthorblockA{\textit{Universität Wien},
        Vienna, Austria\\
        tanja.sarcevic@univie.ac.at\\
        0000-0003-0896-9193\\
    }
\and
    \IEEEauthorblockN{Andreas Rauber}
    \IEEEauthorblockA{\textit{TU Wien},
        Vienna, Austria\\
        andreas.rauber@tuwien.ac.at\\
        0000-0002-9272-6225\\
    }
\and
    \IEEEauthorblockN{Rudolf Mayer}
    \IEEEauthorblockA{\textit{SBA Research},
        Vienna, Austria\\
        rmayer@sba-research.org\\
        0000-0003-0424-5999\\
    }
}

\begin{document}
\maketitle
\thispagestyle{plain}
\pagestyle{plain}

\begin{abstract}
Ensuring data ownership and traceability of unauthorised redistribution are central to safeguarding intellectual property in shared data environments.
Data fingerprinting addresses these challenges by embedding recipient-specific marks into the data, typically via content modifications. 
We propose NCorr-FP, a Neighbourhood-based Correlation-preserving Fingerprinting system for structured tabular data with the main goal of preserving statistical fidelity. 
The method uses local record similarity and density estimation to guide the insertion of fingerprint bits. The embedding logic is then reversed to extract the fingerprint from a potentially modified dataset. 
Extensive experiments confirm its effectiveness, fidelity, utility and robustness.
Results show that fingerprints are virtually imperceptible, with minute Hellinger distances and KL divergences, even at high embeddeding ratios. The system also maintains high data utility for downstream predictive tasks. The method achieves 100\% detection confidence under substantial data deletions and remains robust against adaptive and collusion attacks.
Satisfying all these requirements concurrently on mixed-type datasets highlights the strong applicability of NCorr-FP to real-world data settings.
\end{abstract}

\section{Introduction}\label{sec:introduction}
In an era of widespread data sharing and outsourcing, the ability to assert ownership over structured datasets and trace unauthorised redistribution is increasingly critical for protecting the intellectual property (IP) of such content. 
Data fingerprinting systems address this need by embedding unique, recipient-specific marks into the data, enabling leak attribution and intellectual property protection. 
The existing fingerprinting methods for structured data often fall short in balancing properties and requirements that need to be met simultaneously in real-world settings: high effectiveness (especially in \textit{blind} scenarios where original data is unavailable during detection), robustness to single-user and collusion attacks, mark imperceptibility (data fidelity, i.e. minimal distortion and preservation of statistical properties), data utility and applicability across datasets with mixed-type attributes.

This work introduces NCorr-FP, a Neighbourhood-based Correlation-preserving Fingerprinting system 
designed to achieve high detection confidence while preserving the statistical fidelity and contextual plausibility of structured data. Unlike prior techniques that rely on randomised value manipulations or global assumptions, NCorr-FP exploits local record similarity and attribute correlations to embed fingerprint bits in a way that is both statistically inconspicuous and resilient to a broad range of adversarial threats. 
Our method supports blind detection, requires no access to the original dataset, and integrates with collusion-resilient codes. 

Through extensive evaluation on a benchmark dataset, we demonstrate that NCorr-FP introduces only minimal statistical distortion, with attribute distributions, correlations, and value combinations remaining nearly indistinguishable from the original. 
Additionally, the impact on downstream machine learning tasks is negligible.
Despite these minimal alterations, effectiveness and robustness remain uncompromised: NCorr-FP achieves 100\% fingerprint detection confidence even under extreme conditions such as the removal of up to 80\% of records or 70\% of attributes. Furthermore, the system withstands adaptive, informed attacks—including cluster-flipping strategies, and reliably detects collusion involving up to 25\% of all data recipients.

\begin{figure}[t]
    \centering
    \includegraphics[width=\linewidth]{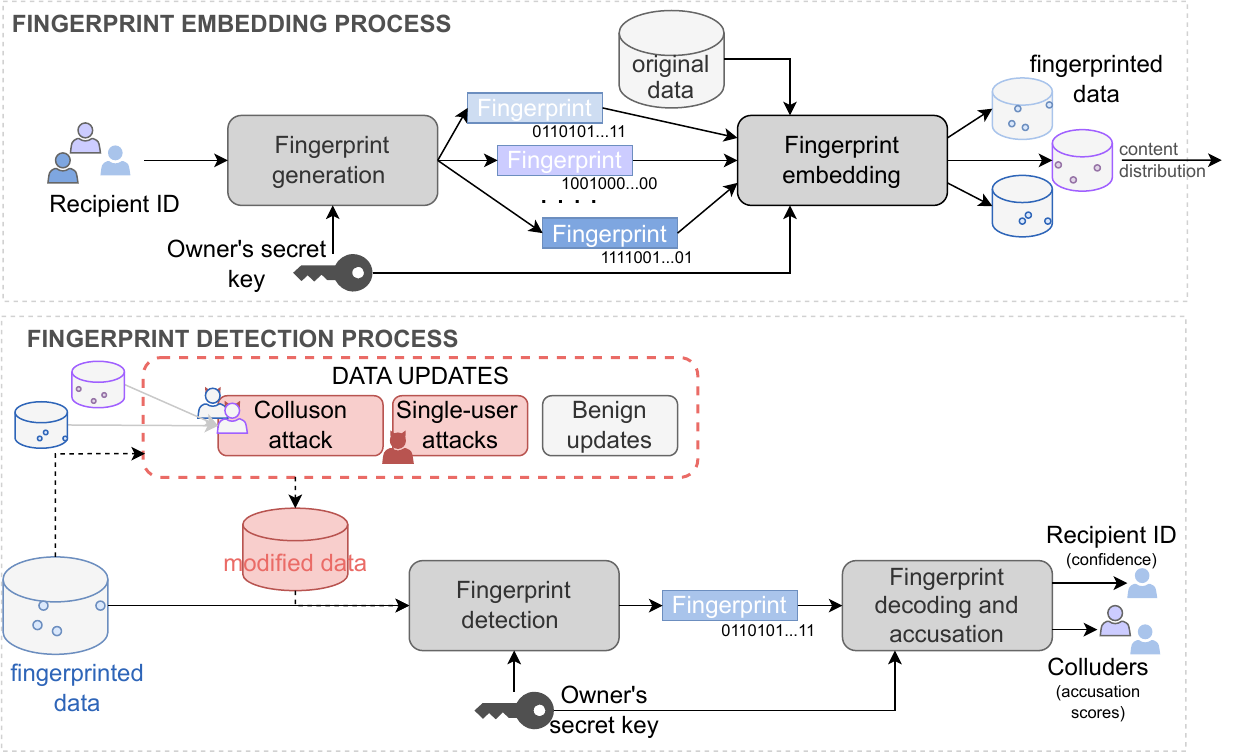}
    \caption{Fingerprinting system: (i) fingerprint embedding process implementing fingerprint generator and fingerprint embedding, and (ii) fingerprint detecting process embedding fingerprint detection and a decoding and accusation mechanism.}
    \label{fig:fp-system}
\end{figure}

In summary, this paper makes the following contributions:
\begin{itemize}
    \item A novel neighbourhood-based fingerprinting algorithm that preserves data fidelity by embedding marks through statistically plausible value substitutions within similar records.
    \item Robustness against single-user and collusion attacks, including adaptive and white-box adversaries, supported by a detailed robustness evaluation.
    \item Comprehensive empirical validation showing superior performance across effectiveness, fidelity, utility, and robustness dimensions.
\end{itemize}

NCorr-FP bridges the gap between robust ownership protection and practical usability in data sharing pipelines.
Our code is available open-source\footnote{\url{https://github.com/sbaresearch/data-fingerprinting}}.

The notation used in this work is summarised in \Cref{tab:notation}. 
The remainder is structured as follows:  
\Cref{sec:system} outlines the system model and \Cref{sec:threat-model} the threat model.
\Cref{sec:related-work} reviews related work.
\Cref{sec:NCorrFP} presents our proposed method, including the fingerprint embedding and detection algorithms.  
\Cref{sec:evaluation-methodology} details the evaluation methodology, covering metrics for effectiveness, fidelity, utility, and robustness. 
\Cref{sec:results} reports and analyses the experimental results. 
Finally, \Cref{sec:conclusions} concludes the paper and discusses future directions.
\begin{table}[ht]
    \centering
    \caption{Notation}
    \label{tab:notation}
    \begin{tabular}{l|l}
        \toprule
        notation & description \\
        \midrule 
        $\mathcal{R}$ & dataset relation \\
        \textit{PK} & primary key of the dataset \\
        $A_i$ & \textit{i}-th attribute in the dataset \\
        \textit{v} & number of attributes in the dataset\\
         \textit{n} & number of records in a dataset \\
         \midrule
         $\mathcal{F}$ & fingerprint bit sequence \\
         \textit{L} & fingerprint length in bits\\
         1/$\gamma$ & fingerprint embedding ratio\\
         $\mathcal{S}$ & pseudo-random sequence generator (PRSG)\\
         $\mathcal{K}$ & (owner's) secret key\\
         $c$ & number of colluders in collusion attack\\
         \midrule
         $\mathcal{N}$ & neighbourhood; a set of records most similar to a target \\ &(reference) record\\
         $d$ & distance metrics for measuring record similarity\\
         $\phi$ & percentile value determining density thresholds\\
         $\mathcal{LD}_\phi$ & low-density or low-fequency region of a distibution\\
         $\mathcal{HD}_\phi$ & hihg-density or high-freqeuncy region of a distribution\\

         $C$ & sets of correlated attributes, $C\subseteq A$ \\
         \bottomrule
    \end{tabular}
\end{table}

\section{System and Requirements}\label{sec:system}
Data fingerprinting system protects ownership of structured datasets by embedding unique, recipient‐specific marks, \textit{fingerprints}, into the data so that any unauthorized redistribution can later be traced back to the leaker.  
Each recipient receives a differently marked copy; a leaked copy may be subjected to benign updates or malicious tampering.
The objective of the fingerprinting process is to embed a verifiable, recipient‐specific fingerprint \(\mathcal{F} = (f_1, \dots, f_L)\) such that:
(i) the fingerprint can be reliably extracted and attributed to the correct recipient,
(ii) the marked data remains useful and statistically similar to the original and (iii) the fingerprint survives partial or intentional data modifications.
Following these objectives, the fingerprinting systems must satisfy the following requriements~\cite{rani_comparative_2022}:
\begin{itemize}
    \item Effectiveness:  
    High-confidence fingerprint detection.
  \item Fidelity:  
    Preservation of statistical properties (marginal distributions, correlations, plausible value combinations).
  \item Utility:
    Minimal impact on a downstream use of data.
  \item Robustness: Resilience to attacks.
  \item Blindness: Ability to extract fingerprints without access to the original data.
\end{itemize}

A fingerprinting system implements two main processes - \textbf{fingerprint embedding} and \textbf{fingerprint detection}, as depicted in \Cref{fig:fp-system}, and the following subprocesses:
\begin{enumerate}
  \item \textbf{Fingerprint generator:}  
    Produces a recipient‐specific bit sequence \(F\) using secret key \(K\) and recipient ID.
  \item \textbf{Embedding algorithm:}  
    Injects \(F\) into the dataset via a content‐aware, density‐ and correlation‐driven scheme (\Cref{subsec:NCorrFP-embedding}).
  \item \textbf{Detection algorithm:}  
    Extracts a candidate fingerprint from a (possibly modified) copy without access to the original (\Cref{subsec:NCorrFP-detection}).
  \item \textbf{Fingerprint decoding and accusation:}  
    Matches the extracted fingerprint against known fingerprints to identify the source or find responsible colluders.
\end{enumerate}
To implement 1) and 4), we incorporate Tardos codes~\cite{tardos_optimal_2008} (T codes) into our system, as they remain one of the most efficient constructions for traitor tracing. 
T-codes are probabilistic fingerprint codes of length 
\begin{equation}\label{eq:tardos-length}
    L = O(c^2 \log(N/\epsilon))
\end{equation}
for a given number of users $N$, collusion size $c$, and error probability $\epsilon$. 
The fingerprint generation and the accusation algorithm follow the proposed construction in Section 2.1. of \cite{tardos_optimal_2008}, we skip the details for brevity but keep the notation.
During accusation, each recipient is assigned an accusation score, $Tscore$.
 We define the accusation threshold $Z$ by the mean accusation score across all recipients $s$, $\mu(Tscore)$, and varible number of standard deviations, $ \sigma(Tscore)$:
\begin{equation}\label{eq:tardos-accusation-threshold}
    Z_x = \mu(Tscore) + x\cdot\sigma(Tscore).
\end{equation}
The recipient $s$ is accused if their score $Tscore_s>Z_x$.

Our modular design permits substituting T-codes with other fingerprinting or watermarking methods without altering the overall system.

\section{Threat Model}\label{sec:threat-model}
As shown in \Cref{fig:fp-system}, we consider the attacker to be a malicious recipient of the fingerprinted data, whose objective is to avoid identification while engaging in unauthorised redistribution. We assume a \emph{white‐box} attacker who:
\begin{itemize}
  \item Knows the fingerprinting algorithm and typical parameter choices (public‐system model~\cite{halder_watermarking_2010}). 
  \item Holds one or more fingerprinted copies (single‐user or colluding recipients).
  \item May modify the fingerprinted data using any combination of deletion, alteration, or injection techniques.
\end{itemize}
We assume the owner’s secret key \(K\) remains unknown to the attacker.

\section{Related Work}\label{sec:related-work}
Fingerprinting and watermarking techniques have been widely studied in the context of digital rights management and applied accross various forms of digital content such as image, video, audio, structured datasets and Machine Learning (ML) models~\cite{barni_information_2023,agarwal_survey_2019}.
In structured data, the focal area of our work, \textit{fingerprinting} refers to recipient-specific watermarking, where each recipient receives a uniquely marked version of the dataset to enable leak tracing. 
In contrast, watermarking typically embeds a common mark to assert ownership.
Foundational work for database watermarking~\cite{agrawal_watermarking_2003} and fingerprinting~\cite{li_fingerprinting_2005} were based on pseudorandom insertion of fingerprint bits into the least-significant bits of data values. 
These works have been extended by optimised embedding strategies, however primarily focussing on numerical data attributes~\cite{liu_block_2005,guo_fingerprinting_2006,lafaye_watermill_2008}. 

Early techniques ensured fidelity by preserving simple statistics such as mean and standard deviation, typically with negligible distortion~\cite{sarcevic_evaluation_2019}. 
More recent approaches incorporate stronger fidelity measures, such as correlation preservation, and address additional requirements like data utility. 
Ji et al.~\cite{ji_curse_2021,ji_towards_2023} propose a post-processing step to any base fingerprinting scheme (the authors demonstrate the effectiveness for Li's scheme~\cite{li_fingerprinting_2005} and block-scheme~\cite{liu_block_2005}) to mitigate disruptions in joint distributions. 
Their method, based on Optimal Mass Transportation, selectively alters unmarked values to restore column-wise correlations and is particularly effective against attackers with prior knowledge of original joint distributions. 
A fingerprinting method for categorical data is proposed in~\cite{sarcevic_correlation-preserving_2020}, which employs a neighbourhood-based strategy to embed marks by selecting plausible alternative values. 
While the approach is designed to preserve local value distributions, it is limited to categorical attributes and does not address applicability to numerical domains. 
Furthermore, the detection phase requires access to the original dataset, which restricts its use in scenarios where blind detection is essential. 

Additional efforts to preserve correlation structures appear in the context of sequential~\cite{ayday_robust_2019,yilmaz_collusion-resilient_2020} and textual data~\cite{perez_gort_semantic-driven_2021}. 
The former cannot be directly applied to structured data because they assume sequential correlation that is not inherent to this data type~\cite{ji_towards_2023}, while the latter cannot generalise to real-world datasets with mixed attribute types. 

Collusion attacks, which arise from malicious collaboration among users, pose a significant threat to fingerprinting systems.
Early fingerprint codes, such as hash-based encodings using secret keys and user IDs~\cite{li_fingerprinting_2005}, are ineffective against collusion.  
To address this, collusion-resistant codes like the Boneh-Shaw (BoSh) code~\cite{boneh_collusion-secure_1998} enforce traceability under the \textit{Marking Assumption}. However, BoSh codes are impractically long. Tardos~\cite{tardos_optimal_2008}, and later Nuida et al.~\cite{nuida_improvement_2009}, introduced probabilistic Tardos codes (T-codes) with reduced lengths suitable for structured data. Group-based schemes~\cite{trappe_anti-collusion_2003,yu_group-oriented_2010} are proposed for content distributed in batches across users, and privacy-preserving extensions~\cite{zhang_collusion-resilient_2024} for an additional privacy guarantee.

\section{NCorr-FP: Neighbourhood-based Correlation-preserving Fingerprinting}\label{sec:NCorrFP}
This section presents our fingerprinting method, NCorr-FP, which is designed to preserve statistical properties and contextual consistency in structured data while embedding recipient-specific marks. 
We begin by describing the fingerprint embedding algorithm, which uses local record similarity and density estimation to guide the insertion of fingerprint bits. 
We then outline the corresponding detection procedure, which reverses the embedding logic to extract the fingerprint from a potentially modified dataset. 
Both procedures operate under the constraint of blind detection, requiring no access to the original data.

\subsection{Fingerprint embedding algorithm}\label{subsec:NCorrFP-embedding} 
The embedding algorithm uses the fingerprint $\mathcal{F}$ as an input and hides its bits $f_i$ in the data by determining the required modification of values at pseudo-random positions. 

This process inevitably introduces changes to the dataset in order to encode verifiable information. 
To minimise the impact of these changes and their perceptibility, each change is sampled from the space of similar records given the current record. 
The goal is to ensure that the new values appear statistically plausible within their local context. 
As a result, the method avoids generating unlikely or inconsistent values and better preserves correlations.

\begin{algorithm}
  \caption{NCorr-FP: Embedding}
  \label{alg:NCorr-FP-embedding}
  \KwIn{dataset $\mathcal{R}$ = $(PK, A_{0}, ..., A_{v-1})$, owner's secret key $\mathcal{K}$, embedding param. $\gamma$, fingerprint $\mathcal{F}$, correlated groups $C$, distance metric $d$, density param. $\phi$}
  \KwOut{fingerprinted dataset $\mathcal{R'}$}
  \ForEach{record $r \in \mathcal{R}$}
  {
    $\mathcal{S}(\mathcal{K}|r.PK)$ = $s_0$, $s_{1}$, $s_{2}$, $s_{3}$\\
    \If{($s_0$ mod $100)/100<1/\gamma$}{
        attribute\_index $i \gets s_1$ mod $v$\\
        fingerprint\_index $l \gets s_2$ mod $L$\\ 
        fingerprint\_bit $f \gets f_l$\\
        mask\_bit $x\gets 0$ if $s_3$ is even; $x\gets 1$ otherwise\\
        mark\_bit $m\gets x \oplus f$\\
        $\mathcal{N}$ = $select\_neighbours(r.A_i,d, C)$\\
        $target\_values\gets \mathcal{N}.A_i$\\
        $\mathcal{H}_p,\mathcal{L}_p\gets density\_areas(target\_values, p)$\\
        \uIf{m = 1}{
            $r'.A_i\gets sample(\mathcal{HD}_\phi)$\\
        }
        \uElse{
            $r'.A_i\gets sample(\mathcal{LD}_\phi)$\\
        }
    }
  }
  \Return{$\mathcal{R'}$}
\end{algorithm}

The embedding algorithm shown in \Cref{alg:NCorr-FP-embedding} traverses the records of the dataset (line 1), initiating in each step a pseudorandom sequence generator (PRSG) $\mathcal{S}$ seeded by the combination of the owner's secret key $\mathcal{K}$ and the primary key of the record $r.PK$ (line 2) (e.g. via a concatenation, denoted with $|$). 
A PRSG of each record $r$ outputs a unique number sequence where the first 4 are relevant outputs:
\begin{enumerate}
    \item $s_0$: determining whether the record is selected 
    \item $s_1$: index of the attribute whose value will be marked
    \item $s_2$: index of a fingerprint bit to embed
    \item $s_3$: \textit{how} the fingerprint will be embedded (mask bit) 
\end{enumerate}
A binary decision on whether the record $r$ is being selected for marking is made in line 3 by using a pseudo-random output $s_0$ weighted by embedding ratio $1/\gamma$
, i.e. every record $r$ is selected for marking with independent probability $1/\gamma$. 
This entails that approximately $n/\gamma$ records of the dataset will be marked in total, hence we refer to $1/\gamma$ as the \textit{embedding ratio} for clarity. 
$s_1$ and $s_2$ are sampled from the range of the total number of attributes $v$ (line 4), and the range of bit-length of the fingerprint, $L$ (line 5), respectively.
$s_3$ determines the binary mask $x$ used to transform the fingerprint bits into the uniformly distributed space. 
The distribution of bits in a fingerprint can be imbalanced, which is more pronounced in shorter fingerprints. 
However, a scenario where one type of bit prevails is not ideal for marking. 
That way, data copies would have contained various amounts of modifications, implying unfair differences in data quality that different recipients obtain and potentially additional vulnerabilities to attacks.
To ensure statistically that a potential bit imbalance does not affect embedding, each selected fingerprint bit $f$ is subjected to an XOR operation with a uniformly sampled mask bit $x$, resulting in the mark bit $m$, which follows a uniform distribution.

The mark bit $m$ is the deciding bit for sampling the new value for the chosen dataset position $(r,A_i)$.
The new value is sampled from the space of records selected according to their similarity to the target record $r$ under the set of parameters $d \cup C$. 
This space is referred to as the \textit{neighbourhood} $\mathcal{N}$ of the target record. 
The intuition behind this is to flip the value into a likely one based on similar records.

\begin{algorithm}
  \KwIn{marking value $r.A_i$, values of correlated attributes $r.C$, $k$, $d$}
  \KwOut{neighbourhood $\mathcal{N}$}
  \If{$A_i$ in $C$}{
    $tree$ = $tree\backslash A_i$
  }

    $\mathcal{N} \gets tree$.query($r.C$, $k$)\\
    $max\_d_N \gets$ max\_distance($r.C$, $\mathcal{N}$)\\
    $\mathcal{N}$ = $tree$.query\_radius($r.C$, $max\_d_N$)\\
  \Return{$\mathcal{N}$}
  \caption{$select\_neighbours$ sub-procedure}
  \label{alg:NCorr-FP-neighbourhood}
\end{algorithm}
The subroutine $select\_neighbours$ (\Cref{alg:NCorr-FP-neighbourhood}) leverages a \textit{k}-nearest neighbour (\textit{k}NN) imputation strategy. 
\textit{k}NN selects the \textit{k} most similar records to the observed record based on a distance metric $d$ and a list of correlated groups of attributes $C=(C_1, C_2, ...)$ where $\forall i C_i\subseteq A$.

We introduce a partitioning of the attribute space into correlated groups of attributes $C = (C_1, C_2, \dots)$, where each $C_i \subseteq A$ contains attributes that are mutually and transitively correlated. 
 Embedding modifications within a correlated group helps preserve the statistical properties of the dataset. Changes made to an attribute $A_j$ in group $C_i$ are informed by other attributes in the same group that share high mutual dependence. 
 This local dependency structure allows us to estimate realistic value intervals from the neighbourhood and to constrain the embedding noise accordingly. 

The correlated sets in $C$ are computed once prior to the embedding based on a correlation metric $corr$ and inclusion threshold $\tau_c$. 
Formally, $C$ may be defined as a correlation graph $C=G(A, E)$ where the nodes are the attributes in A and E, a set of edges. An edge exists between $A_i$ and $A_j$ if $|corr(A_i, A_j)|>\tau_c$, i.e. $E=\{(A_i, A_j)| |corr(A_i, A_j)|> \tau_c \}$.
The groups of mutually correlated attributes correspond to the connected components $C_1,C_2,\dots,C_g$ in $G$ where $A = \bigcup_{i=1}^{g} C_i, \quad C_i \cap C_j = \emptyset \text{ for } i \neq j$. Each group $C_i$ contains attributes that are transitively correlated, meaning that if $A_i$ is correlated to $A_j$ and $A_j$ is correlated to $A_k$, then $A_i$ and $A_k$ belong to the same group even if $|corr(A_i, A_k)|\leq\tau_c$.

The $select\_neighbours()$ procedure can be adapted via usual practices for designing \textit{k}NN for specific data and problem. 
This includes choosing a distance metric (e.g. Euclidean Distance, L1 norm, cosine similarity, etc.), improving efficiency via data structures specifically designed for efficient search (in this work we used a Ball Tree structure, denoted as $tree$ in \Cref{alg:NCorr-FP-neighbourhood}, but other structures can be used as well, e.g. a KD-Tree) and hyperparameter \textit{k}. 
We analyse the impact of \textit{k} on the overall scheme effectiveness in \Cref{sec:results}. 
The algorithm may be non-deterministic in cases where $x$ records with the same distance (i.e. similarity) to the target record compete for $k < x$ places in the neighbourhood. 
To tackle this, in the $select\_neighbours$ subprocedure we expand the neighbourhood with all candidates that tie in the distance, hence we treat $k$ as a \textit{minimal} neighbourhood size (lines 5-6), and output such constructed neighbourhood $\mathcal{N}$ in line 7.  

\begin{figure}[t]
     \centering
     \subfloat[\label{fig:demo-continuous-embedding}Embedding (original data)]{%
       \includegraphics[height=70pt]{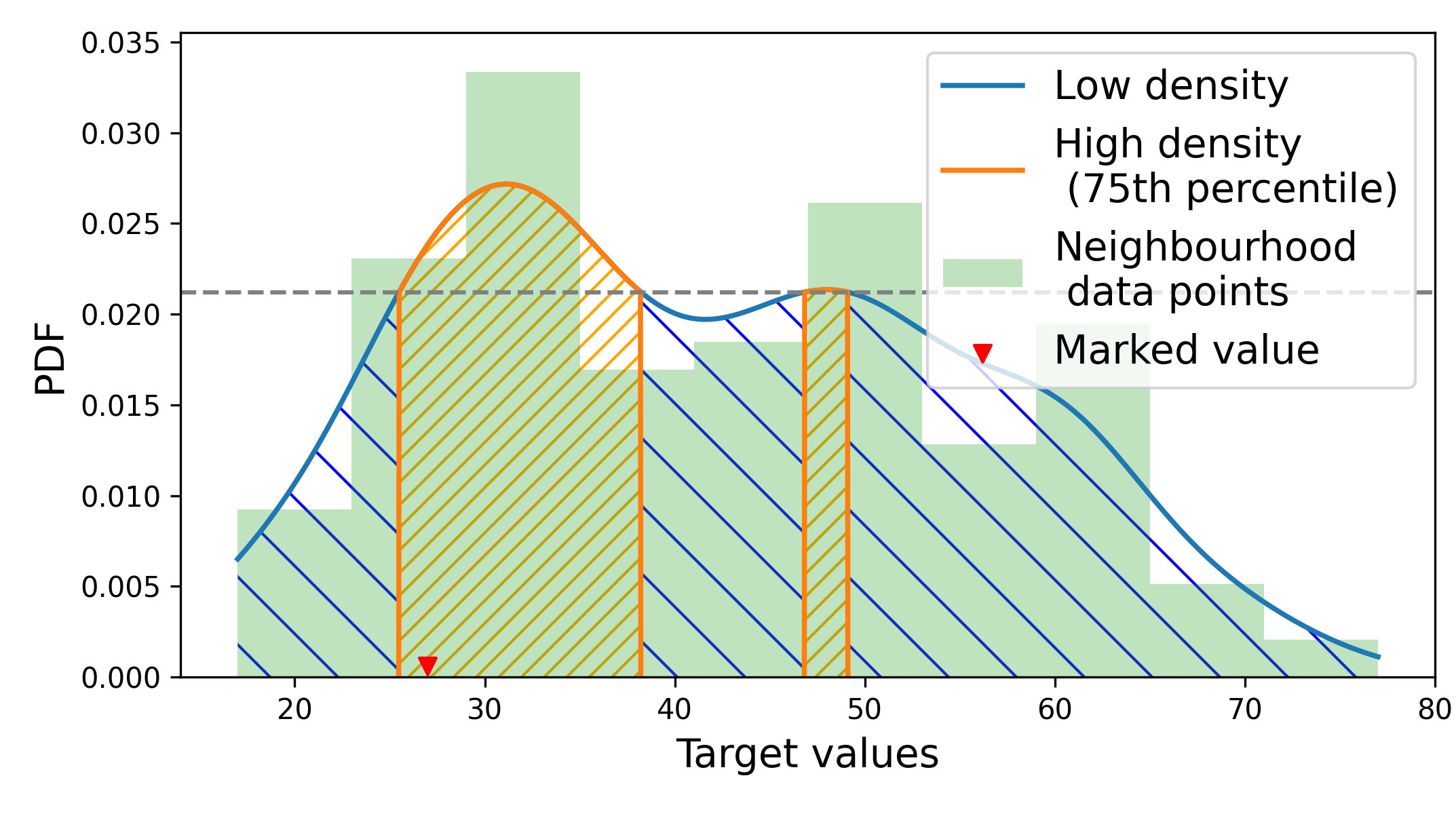}}
    \hfill
  \subfloat[\label{fig:demo-continuous-detection}Detection (fingerprinted)]{%
        \includegraphics[height=70pt]{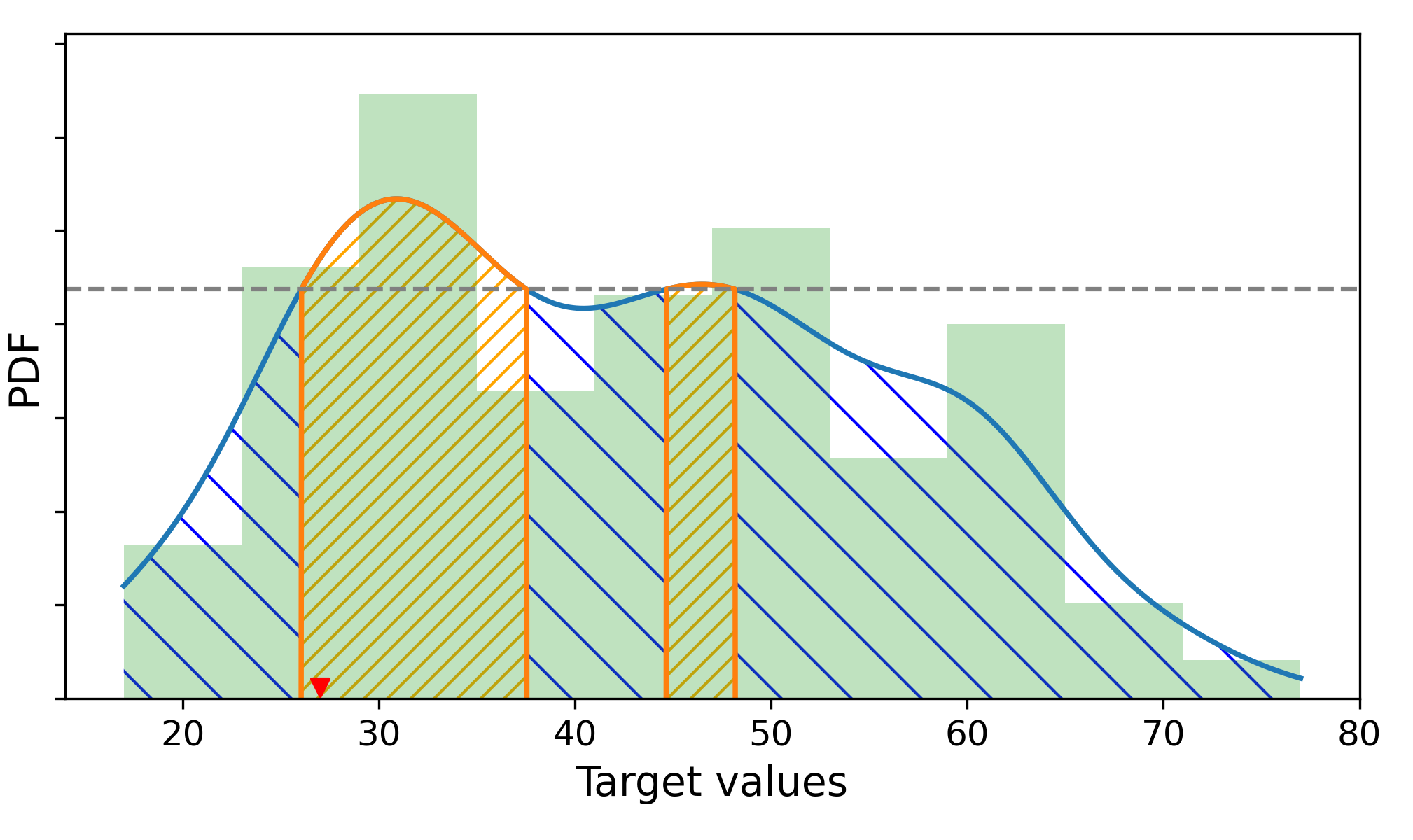}}
         \caption{NCorr-FP demonstration for continuous attributes: during embedding in a), PDF of target values from a neighbourhood $\mathcal{N}$ for sampling the new value $r'.A_i$ is divided into low- (blue) and high-density areas (orange) using a density percentile $\phi=75\%$. Here we exemplify sampling from a high-density area when $m=1$. The detection process in b) regenerates the PDF of target values from the fingerprinted data. Observe that the distributions in a) and b) are not identical -- this is due to the changes introduced by the fingerprint marks. They are, however, very similar. Therefore, the low- and high-density division of b) closely matches that of the original data in a), and the observed value is correctly classified into the high-density area in this example, therefore, the (correctly) detected mark bit is $m=1$.}
     \label{fig:demo-continuous}
 \end{figure}

$target\_values$ are the values of target attribute $A_i$ that occur within the neighbourhood $\mathcal{N}$ (line 10 of \Cref{alg:NCorr-FP-embedding}). 
In lines 12-15, $sample()$ function performs sampling from the set of $target\_values$ and has strictly two outcomes based on the mark bit $m$:
\begin{itemize}
    \item $m=0$: the new value $r'.A_i$ is randomly sampled from the low-density areas of the $target\_values$ distribution; $\mathcal{LD}_\phi$
    \item $m=1$: the new value $r'.A_i$ is randomly sampled from the high-density areas of $target\_values$ distribution; $\mathcal{HD}_\phi$
\end{itemize}
The procedure $density\_areas()$ (line 11) outputs the high- and low-density areas and is adapted to the data type of the $A_i$.
For continuous attributes, we first estimate the probability density function $\hat{f}$ using the Gaussian Kernel Density Estimation method (Gaussian KDE) to obtain the sampling space:
\begin{equation}
\hat{f}(x) = \frac{1}{n h \sqrt{2\pi}} \sum_{i=1}^{n} \exp\left(-\frac{(x - x_i)^2}{2h^2}\right).
\end{equation}
where $\hat{f}$ is the estimated density at point $x$, $n$ is the number of data points, $h$ is the smoothing parameter and $x_i\in target\_values$ are the observed data points.
The boundary between the low and dense distribution areas is determined by the predefined percentile $\phi\in(0,1)$. 
The density threshold $\tau_\phi$ is defined as the value satisfying:
\begin{equation}\label{eq:density_threshold}
    P(\hat{f}(X) \leq \tau_\phi) = \phi
\end{equation}
where $X$ is a random variable following the estimated density $\hat{f}(x)$. This means that a proportion $\phi$ of the estimated density values lie in the low-density region.
Thus, we classify the density space into a low-density region $\mathcal{LD}_\phi$ and high-density region $\mathcal{HD}_\phi$ such that:
\begin{align}\label{eq:density_regions}
    \mathcal{LD}_\phi &= \{ x \in \text{target\_values} \mid \hat{f}(x) \leq \tau_\phi \} \\
    \mathcal{HD}_\phi &= \{ x \in \text{target\_values} \mid \hat{f}(x) > \tau_\phi \}
\end{align}
Note that any probability density estimation may substitute the Gaussian KDE we used in this work. 
We demonstrate in \Cref{fig:demo-continuous-embedding} how the estimated PDF may look for one record (i.e. one iteration of the \textit{foreach} loop in line 1 of \Cref{alg:NCorr-FP-embedding}) and how it gets divided into high- (orange) and low-density (blue) regions based on $\phi$.

\begin{figure}[t]
     \centering
     \subfloat[\label{fig:demo-categorical-embedding}Embedding (original data))]{%
       \includegraphics[height=70pt]{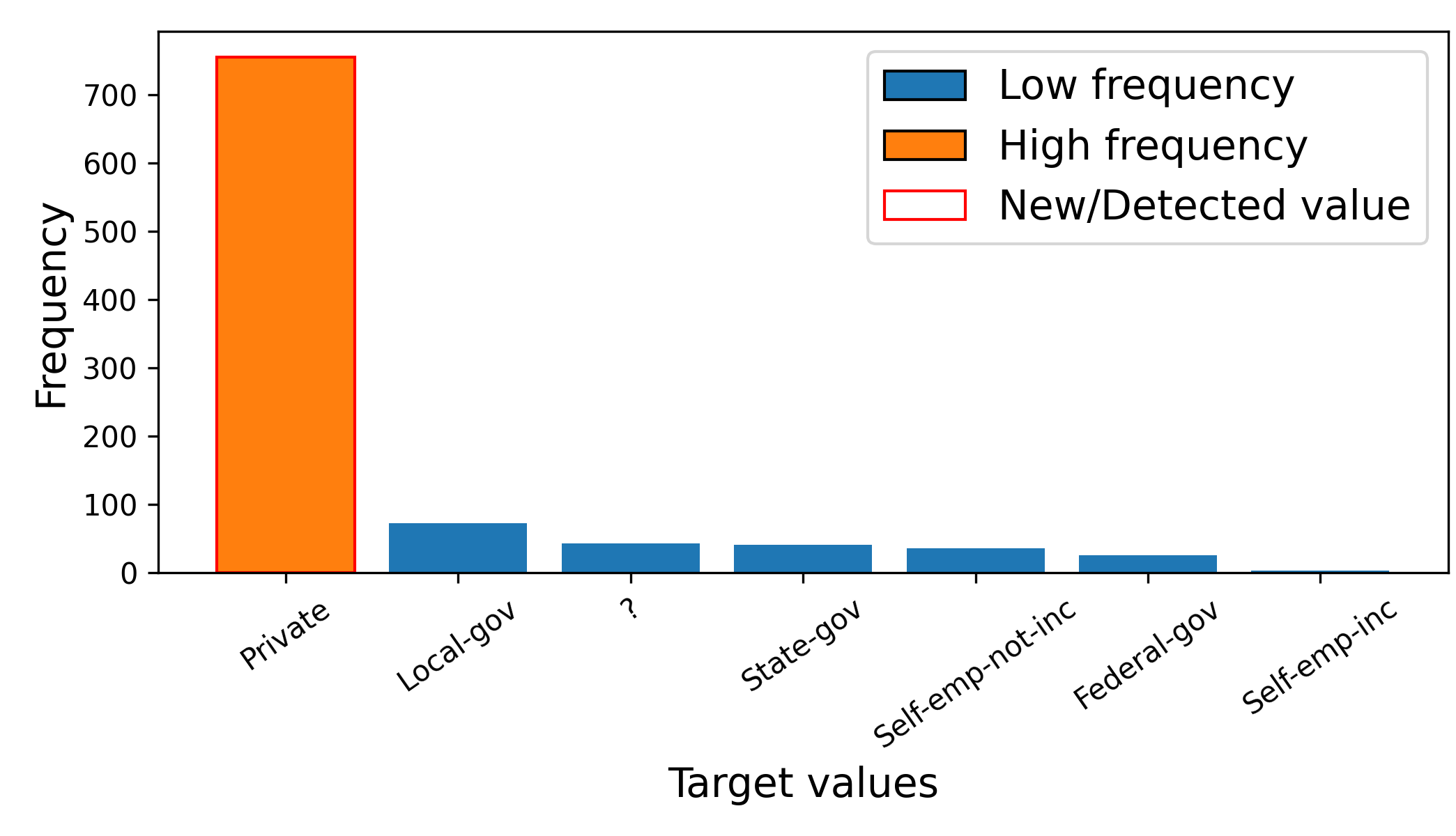}}
    \hfill
    \subfloat[\label{fig:demo-categorical-detection}Detection (Fingerprinted)]{%
       \includegraphics[height=70pt]{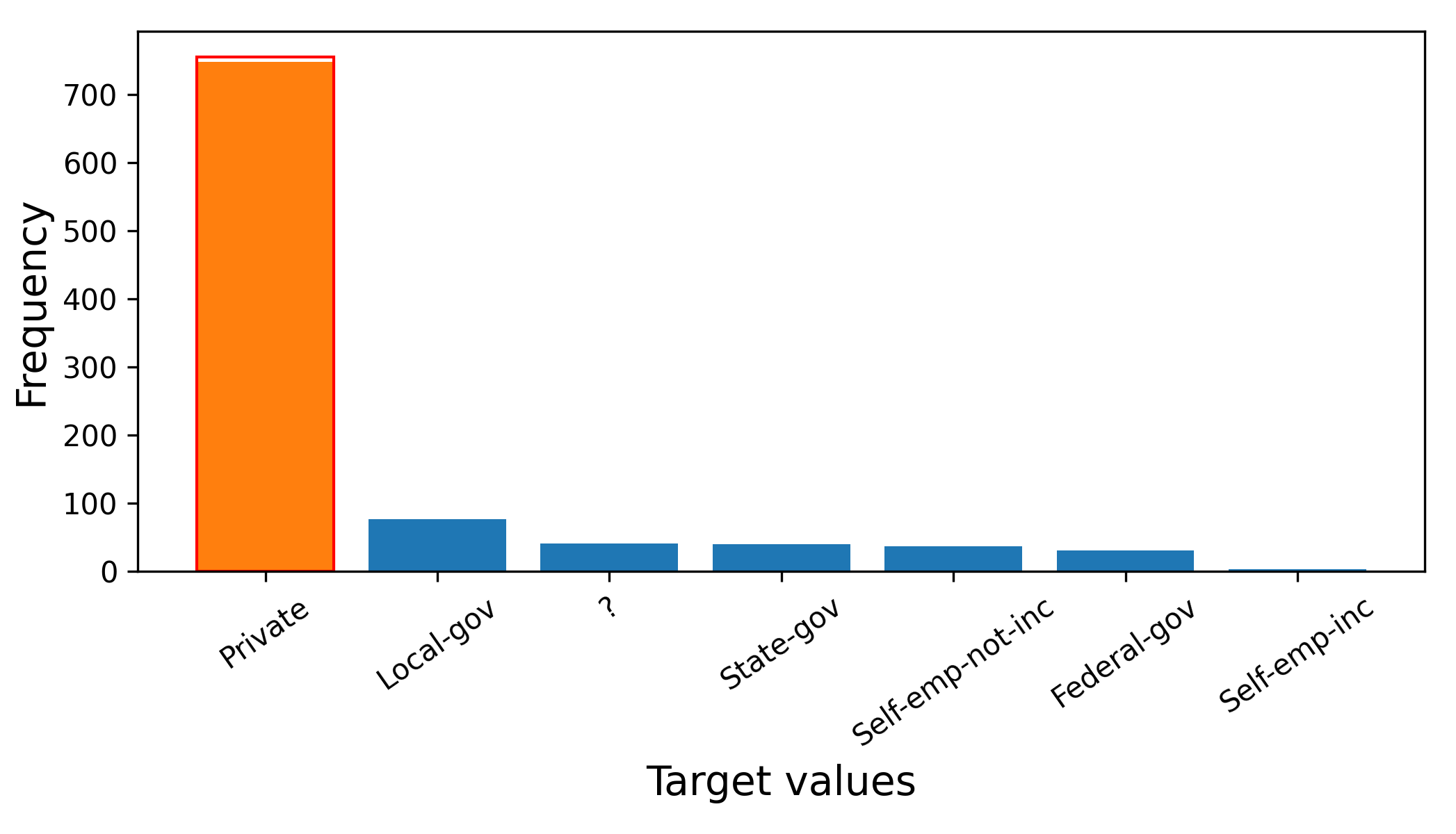}}
     \caption{NCorr-FP demonstration for categorical attributes: in a) the frequencies of the target values from the neighbourhood $\mathcal{N}$ are sorted and grouped: the high-frequency group $\mathcal{HD}_\phi$ contains the least amount of most frequent values such its cumulative frequency is at least $1-\phi=25\%$ of the cumulative frequency of all target values (orange). The rest is the low-frequency group $\mathcal{LD}_\phi$ (blue). This example shows the case when $m=1$, i.e. the new (fingerprinted) value is chosen from the high-frequency group. In b) the detection algorithm generates the frequencies from the fingerprinted data which are highly similar to the original (but not necessarily identical). It recognises that the observed value is in $\mathcal{HD}'_\phi$, hence the (correctly) detected mark bit is $m=1$.}
     \label{fig:demo-categorical}
 \end{figure}

For categorical attributes, the same principle is applied to separate the high- and low-density areas according to value frequencies in $target\_values$.
This is demonstrated for one record (one iteration in the \textit{foreach} loop in line 1 of \Cref{alg:NCorr-FP-embedding}) in \Cref{fig:demo-categorical-embedding}.
Let $\mathcal{A}=\{a_1, \dots, a_{|unique(target\_values)|}\}$ be the finite set of unique target values of a categorical attribute $A_c$ sorted in ascending order by frequency. 
Furthermore, let $freq:S\rightarrow \mathbb{N}$ be a frequency function that assigns each categorical value $a_i$ its cardinality in the $target\_values$ set and $\phi$ be the predefined percentile for separating low- and high-density groups. 
We define the low-density group $\mathcal{LD}_\phi$ and the high-density group $\mathcal{HD}_\phi$ of a categorical attribute such that $\mathcal{LD}_\phi$ is constructed by accumulating the lowest-frequency values until their collective frequency sum reaches (but does not exceed) $\phi$:
\begin{equation}
    \sum_{a \in \mathcal{LD}_\phi} freq(a) \leq \phi \sum_{a \in \mathcal{A}} freq(a).
\end{equation}
The algorithm ensures that both low- and high-density groups contain at least one element, i.e. $|\mathcal{HD}_\phi| \geq 1$ and $|\mathcal{LD}_\phi| \geq 1$.
When $target\_values$ is uniform (i.e. $|unique(target\_values)|=1$), that constraint cannot be satisfied. 
In this case, the choice can be to (i) proceed with sampling the new value from the uniform distribution or (ii) skip the embedding for that value. Both might lead to detection errors: for (i), the detection has a 50\% chance to blindly guess the mark bit $m$ that was supposed to be embedded at that position (but was not properly embedded because there was no choice in sampling), while for (ii), the detection might be affected by the changes in the neighbourhood distributions after the data is fingerprinted, not properly skip this value if it does not appear uniform during the detection and detect an erroneous bit.
We discuss challenges merging from these potential detection errors further in \Cref{subsec:NCorrFP-detection}.

With the described procedures for separating the distribution into $\mathcal{LD}_\phi$ and $\mathcal{HD}_\phi$, we create a binary setting for sampling both continuous and categorical data. 
This binary decision represents the encoded information, i.e. the mark bit $m$ which is a function of a fingerprint bit $f$ (for $m=1$, the sampling is exemplified in \Cref{fig:demo-continuous,fig:demo-categorical}; when $m=0$, the algorithm samples from the blue regions $\mathcal{LD}_\phi$). 
Note that each fingerprint bit $f_i$ is embedded multiple times across the dataset. $\omega_i$ denotes the number of embeddings (the \textit{redundancy}) of the fingerprint bit $f_i$. 
Since choosing a fingerprint index is an independent random draw, 
\begin{equation}\label{eq:omega}
    \omega_1\approx\omega_2\approx ...\approx\omega_L = \overline{\omega}\approx\frac{n}{L\gamma}
\end{equation}
 where $\overline{\omega}$ is the mean redundancy of a single fingerprint bit.

\subsection{Fingerprint detection algorithm}\label{subsec:NCorrFP-detection}
The detection algorithm reconstructs the fingerprint sequence one bit at a time by searching and retrieving their embeddings in the data. 
Since one fingerprint bit $f_i$ might be embedded in multiple values, the extracted bits are counted for each bit position. The final bits are decided by majority voting from these counts.
To make a binary decision based on a data value, whether the embedded mark was 1 or 0, the detection mimics the embedding: it needs to be able to classify the value into either low- or high-density distribution area (because the observed value has been sampled from either low- or high-density area during embedding). 
For that, the neighbourhoods also need to be reconstructed.
Note that the detection algorithm does not recover the original data value, only the embedded information based on the marked values.

\begin{algorithm}
  \KwIn{fingerprinted database $\mathcal{R'}=(PK, A_0, ..., A_{v-1})$, owner's secret key $\mathcal{K}$, embedding param. $\gamma$, fp. length $L$, correlated groups $C$, distance metric $d$, density param. $\phi$}
  \KwOut{fingerprint $\mathcal{F}$}
  fingerprint template $\mathcal{F}\gets(f_1,...,f_{L})\gets(?,...,?)$  \\
  $count[j][0]=count[j][1]\gets 0$ \textbf{for} $j=1$ to $L$\\ 
  \ForEach{record $r \in R'$}
  {
    $\mathcal{S}(\mathcal{K}|r.PK)$ = $s_{0}$, $s_{1}$, $s_{2}$, $s_{3}$ \\
    \If{($s_0$ mod $100)/100<1/\gamma$}{
        attribute\_index $i\gets s_1$ mod $v$\\
        fingerprint\_index $l\gets s_2$ mod $L$\\
        mask\_bit $x\gets 0$ if $s_3$ is even; $x\gets 1$ otherwise\\
        $\mathcal{N}' \gets select\_neighbours(r.A_i,r.C)$ \\
        $target\_values' \gets \mathcal{N}'.A_i$\\
        $\mathcal{HD}_\phi', \mathcal{LD}_phi' \gets density\_areas(target\_values',\phi)$\\
        \uIf{$r.A_i \in \mathcal{HD}_\phi'$}{
            mark\_bit $m \gets 1$
        }
        \uElse{
            mark\_bit $m\gets 0$
        }
        fingerprint\_bit $f\gets m \oplus x$\\
        $count[l][f]++$ \\
    }
  }
  $\mathcal{F}\gets majority\_voting(count[1], ..., count[L])$
  \caption{NCorr-FP: Detection}
  \label{alg:NCorr-FP-detection}
\end{algorithm}
The detection algorithm reverses the embedding process, cf. \Cref{alg:NCorr-FP-detection}; from the fingerprinted data, it extracts the fingerprint sequence using the owner's secret key $\mathcal{K}$ as an input.
The fingerprint parameters including $\gamma$, $L$, $C$ and $d$ (cf. \Cref{tab:notation}) need to be the same as in the embedding process to correctly detect the fingerprint.
The output of the detection process is the extracted fingerprint bit sequence $\mathcal{F}$, which is initialised in line 1 as a template of $L$ unknown bit-values denoted as $?$.
The unknown values are to be replaced with the actual bit values based on voting throughout the algorithm. 
The voting system is initialised in line 2, where each of the $L$ fingerprint bits collects the votes for its value being 0 or 1 in the variable $count[j][0]$ and $count[j][1]$, respectively.

Following the steps from the embedding, the detection traverses the data records (line 3), this time collecting the information embedded in the pseudo-random locations determined by the PRSG. 
Note that PRSG generates the same sequence $s_0, s_1, s_2, s_3$ as in the embedding algorithm, as it is initialised with the same seed in each iteration $\mathcal{K}|r.PK$ (line 4). 
Therefore, the record selection (line 5), the attribute (line 6), the fingerprint bit (line 7) and the mask (line 8) shall be the same in embedding and detection for each iteration. 

Once the location $(r, A_i)$ of the embedded mark is known, the algorithm selects the neighbouring records in the same fashion as the embedding algorithm via \Cref{alg:NCorr-FP-neighbourhood}) and obtains the distribution of the target values (lines 9-10).
The distributions are constructed as in the embedding (cf. \Cref{subsec:NCorrFP-embedding}) via Gaussian KDE for continuous and frequency sorting for categorical attributes, as well as divided into low- and high-density regions using the same percentile $\phi$.
Note that the neighbourhood $\mathcal{N'}$ of the record $r'$ might be different in the detection phase compared to the neighbourhood $\mathcal{N}$ during the embedding phase due to the fingerprint-induced modifications in the data $\mathcal{R}'$.
We showcase this for one record $r$ where the target (marked) value is continuous: by comparing \Cref{fig:demo-continuous-embedding} and \Cref{fig:demo-continuous-detection} the minor shift in distribution can be observed. Similarly, the minor shift can be observed between \Cref{fig:demo-categorical-embedding} and \Cref{fig:demo-categorical-detection} when the target (marked) value is categorical.
The resulting distribution regions $\mathcal{HD}_p'$ and $\mathcal{LD}_p'$ might also differ slightly compared to the corresponding embedding iteration. However, with the recommended fingerprinting parameter settings (recommendations discussed in \Cref{subsec:results-summary}), these shifts will not cause significant errors in the detection; this is because (i) the modifications due to fingerprint embedding are minute relative to the neighbourhood size and (ii) each fingerprint bit is embedded multiple times across different attributes, hence majority voting will be able to mitigate these minor detection errors.
We show this empirically in \Cref{subsec:results-effectiveness}. 

To extract the embedded binary information, the detection algorithm determines whether the observed target value $r'.A_i$ falls within high- or low-density areas and assigns the mark bit $m$ accordingly (lines 12-15). 
The value of the fingerprint bit $f$ is then calculated from the mark bit and mask bit as $f=m \oplus x$ (line 16)~\footnote{Note: $x=y\oplus z \implies y=x\oplus z$} .
The voting system $count$ is updated for the $f_l$ with the extracted bit value in line 17.
When all votes are collected, the final fingerprint bit-sequence is decided based on majority voting (line 19). 
In case where some bits do not get any vote or votes end up in a tie, they retain the undecided value $?$ from the fingerprint template.  
The likelihood of such undecided bits increases when the average embedding redundancy $\overline{\omega}$ is low, as fewer observations are available for each bit.
While undecided values reduce the total number of bits that can contribute positively to the fingerprint detection, thus potentially lowering the overall detection confidence, they are also neutral in the sense that they do not introduce erroneous evidence. 
In that regard, it is preferable to leave a bit undecided rather than risk falsely attributing it to an incorrect value. This design choice helps balance sensitivity with robustness against false accusations.

\section{Evaluation Methodology}\label{sec:evaluation-methodology}
We evaluate our method on the train set of Adult Census dataset from UCI ML repository, as this dataset is often used in the literature~\cite{ji_towards_2023,zhang_collusion-resilient_2024}, and it has a number of properties interesting to explore in the context of our proposed method: (i) both numerical and categorical features, (ii) correlated features, (iii) missing data values.
The dataset contains 14 attributes (plus one target attribute for an ML classification task), and 32,560 records.
We analyse the impact of different parameters on the system requirements using the following values:
\begin{itemize}
    \item Neighbourhood size \textit{k}: [30, 300, 450]
    \item Embedding ratio ($1/\gamma$): [1/2, 1/4, 1/8, 1/16, 1/32] 
    \item Fingerprint length \textit{L}: [128, 256, 512]
\end{itemize}
To control for variance, each evaluation run is repeated 10 times with different random seeds; we report their averages and standard deviations.
For each of the requirements of the fingerprinting system defined in \Cref{sec:system}, we select fitting evaluation measures.

\subsection{Effectiveness}\label{subsec:evaluation-methodology-effectiveness} 
\textbf{Vote Error Rate (VER)}: from the slight shifts in the distributions due to the embedded fingerprint, the bits might, in some iterations, be extracted wrongly. 
    Although the voting system is in place to overcome the influence of up to a certain number of these wrong votes, we quantify their occurrence under different parameter settings. VER is the ratio between the count of wrong votes across all fingerprint bits and the total number of votes.

\textbf{Detection Confidence (DC)}: the bit-wise match rate between the detected and the correct fingerprint. Compared to the VER, DC depicts more closely the actual extraction success for the fingerprint. Normally, $DC=100\%$ is expected to identify the data recipient, i.e. all bits are correctly extracted. However, a high percentage associated with a fingerprint of one recipient might be considered a sufficient indicator to identify the correct recipient, especially if the false accusation confidence (cf. the bullet point below) is low.

\textbf{False Accusation Confidence (FAC)}: a bit-wise match between the detected fingerprint and the fingerprints of the other recipients (fingerprints not embedded into the observed data copy). In the correct setting, the expected value of this measure is around 50\% for random guessing the bit-sequence and always significantly lower than the DC.

\subsection{Fidelity}\label{subsec:evaluation-methodology-fidelity}
We can group them into (i) dataset statistics, which measure the overall dataset change, (ii) univariate statistics, which measure attribute statistics and (iii) multivariate statistics, which measure the correlation between data attributes.

\subsubsection{Dataset statistics}
 \textbf{Data accuracy}\footnote{Not to be confused with classification accuracy from utility analysis.} $Acc(\mathcal{R}')_{rel}=1-(\mathcal{R}\oplus\mathcal{R}')/|\mathcal{R}|$ is the absolute or relative number of data values that remained unchanged in the presence of a fingerprint.
    The lower limit can be expressed through $\gamma$:
    \begin{equation}\label{eq:data-accuracy}
    Acc(\mathcal{R}')_{rel} \gtrapprox 1 - \frac{1}{\nu\gamma} 
    \end{equation}
    i.e. accuracy is bounded by the number of values selected for marking throughout the embedding process. 

\subsubsection{Univariate statistics}
For qualitative evaluation, we visually compare \textbf{histograms} of fingerprinted and original data. 
To quantify the similarities between the attribute distributions of original and fingerprinted data, we use \textbf{Hellinger distance} and \textbf{ Kullback-Leibler (KL) Divergence}.

\subsubsection{Multivariate and correlation statistics}
Qualitative analysis is based on \textbf{pairwise histogram comparison} between the original and fingerprinted data. We quantify the changes in data correlations using correlation measurements appropriate for data types involved: \textbf{Pearson's correlation coefficient} for linear correlation between two ratio variables, \textbf{Cramer's V} for association between two nominal variables and \textbf{Eta-squared} $\eta^2$ for the correlation between a categorical and a ratio variable.

\subsection{Utility}\label{subsec:evaluation-methodology-utility}
We adopted a predictive modelling setup for the utility evaluation of fingerprinted data. 
Specifically, we assessed how the presence of fingerprints affects the performance of machine learning models on a classification task. 
Utility is quantified by measuring the \textbf{change in classification accuracy} between models trained on the original dataset and those trained on its fingerprinted counterpart.
We employ four commonly used classifiers representing a diverse range of learning paradigms: Logistic Regression, Multi-layer Perceptron (MLP), Random Forest, and XGBoost and use a consistent hyperparameter configuration obtained by optimising the performance on the original (non-fingerprinted) data\footnote{ Logistic Regression: \{solver: newton-cg, C: 10\},
  MLP: \{solver: adam, learning\_rate: invscaling, hidden\_layer\_sizes: (50,), alpha: 0.001, activation: relu\},
   Random Forest: \{n\_estimators: 120, criterion: entropy\},
    XGBoost: \{subsample: 0.5, reg\_alpha: 0.01, n\_estimators: 160, max\_depth: 10, learning\_rate: 0.009, colsample\_bytree: 0.9\}}. 
For each classifier, we performed 10-fold cross-validation to ensure the robustness and generalizability of the performance estimates.

\subsection{Robustness}\label{subsec:evaluation-methodology-robustness} 
\subsubsection{Single-user attacks}\label{subsubsec:evaluation-methodology-robustness-single-user} 
Fingerprints may be subject to malicious attacks and benign updates on the dataset, as identified e.g. by Rani and Halder~\cite{rani_comparative_2022}.
Several attacks, such as taking horizontal and vertical subsets and value-flipping, implement the types of modifications that may affect fingerprint extraction:
\begin{itemize}
    \item Horizontal subsetting attack: removing a percentage of data records
    \item Vertical subsetting attack: removing a percentage of data columns
    \item Flipping attack: flipping a percentage of data values, e.g. to another value from the attribute domain.
\end{itemize}
Besides randomised approaches to these modifications, we take the perspective of a strong and knowledgeable attacker and tailor the modifications to our specific fingerprinting scheme. 
We recognise that the heuristic nature of the embedding algorithm, which lies in marking values according to their similarity, may open a potential attack vector.
To follow the public-system requirement~\cite{halder_watermarking_2010}, we assume that the attacker knows the algorithmic steps of the embedding and extraction processes, as well as fingerprint parameter selection guidelines (hence, may guess or closely approximate the actual values), i.e. white-box access. 
To that end, we propose a cluster-flipping attack.

\textbf{Cluster-flipping attack} is an attack tailored for our scheme where the attacker uses the white-box capabilities. 
The attacker runs an embedding algorithm on their fingerprinted copy (the available resource similar to the original data) using their own secret key. 
They use this process as a proxy to find the most influential records in neighbourhood construction. 
Frequency of a single record in all caluated neighbourhoods indicates its higher influence on the overall fingerprinting process, hence we hypothesise that disrupting those frequent records has a higher influence on the resulting fingerprint detection.
The attacker then selects a percentage of the most influential records and performs a flipping attack within this group. 
We explore both the case when the attacker closely approximates the fingerprinting parameters, and when the attacker uses the exact parameters used for the original embedding. Note that the attacker's embedding simulation is the approximation of the actual embedding: the embedding cannot be fully replicated without access to the owner's key $\mathcal{K}$, which is kept secret.  
   
To measure the robustness of a fingerprint against attacks, we need to consider both the success of the attacker removing the fingerprint and the associated costs the modifications incur, e.g. by measuring the remaining data fidelity.  
For measuring the attack success, we utilise the \textbf{Detection Confidence (DC)}, i.e. the bit-wise match rate between the detected and correct fingerprint as introduced in \Cref{subsec:evaluation-methodology-effectiveness}. 
For the  \textbf{Attacker's loss} or costs, we measure the reduction of the fidelity (cf. \Cref{subsec:evaluation-methodology-fidelity}).

\subsubsection{Collusion attacks}
A collusion attack is carried out by the cooperation of two or more recipients of data copies, each of which comprises individual fingerprints. 
Collusion attacks can be differentiated based on the aggregation strategy of the collaborators, including, but not limited to:
\begin{enumerate}
    \item Averaging attack: different dataset copies are averaged to smooth out differences and remove unique identifiers
    \item Substitution attack: common values across multiple copies are retained while the differences are substituted by a new value from an attribute domain that is contained in none of the colluders' data (if the domain is large enough) 
\end{enumerate}
Collusion attacks may be extended by additionally flipping some of the agreeing values to increase the chance of coincidental deletion of fingerprint bit embeddings (\textit{substitution and flip}).

The goal of our collusion analysis is to assess the ability of the fingerprinting system to reliably identify guilty parties while minimising the risk of falsely accusing innocent recipients.
To quantify robustness against collusion, we adopt three evaluation measures that jointly capture the effectiveness and reliability of the fingerprint detection process:
\begin{itemize}
    \item \textbf{Precision}: the number of correctly identified colluders divided by the total number of accusations. This rate is 1.0 when all accused recipients are indeed the colluders.
    \item \textbf{False Accusation Rate} (false discovery rate): the number of innocent recipients accused relative to the total accusations (i.e. 1-\textit{precision}). This measure is ideally 0 when there are no innocent recipients accused. Note that this is different from the effectiveness measure FAC described in \Cref{subsec:evaluation-methodology-effectiveness}. 
    \item \textbf{Recall}: the number of colluders accused relative to the total number of colluding partners. Recall is ideally 1.0 when all colluders are identified.
\end{itemize}
The ideal outcome is high precision and high recall, meaning that all actual colluders are correctly identified, and no innocent recipients are accused. 

\section{Evaluation results}\label{sec:results}
\subsection{Effectiveness}\label{subsec:results-effectiveness}
\begin{figure}[t]
    \centering
    \includegraphics[width=\linewidth]{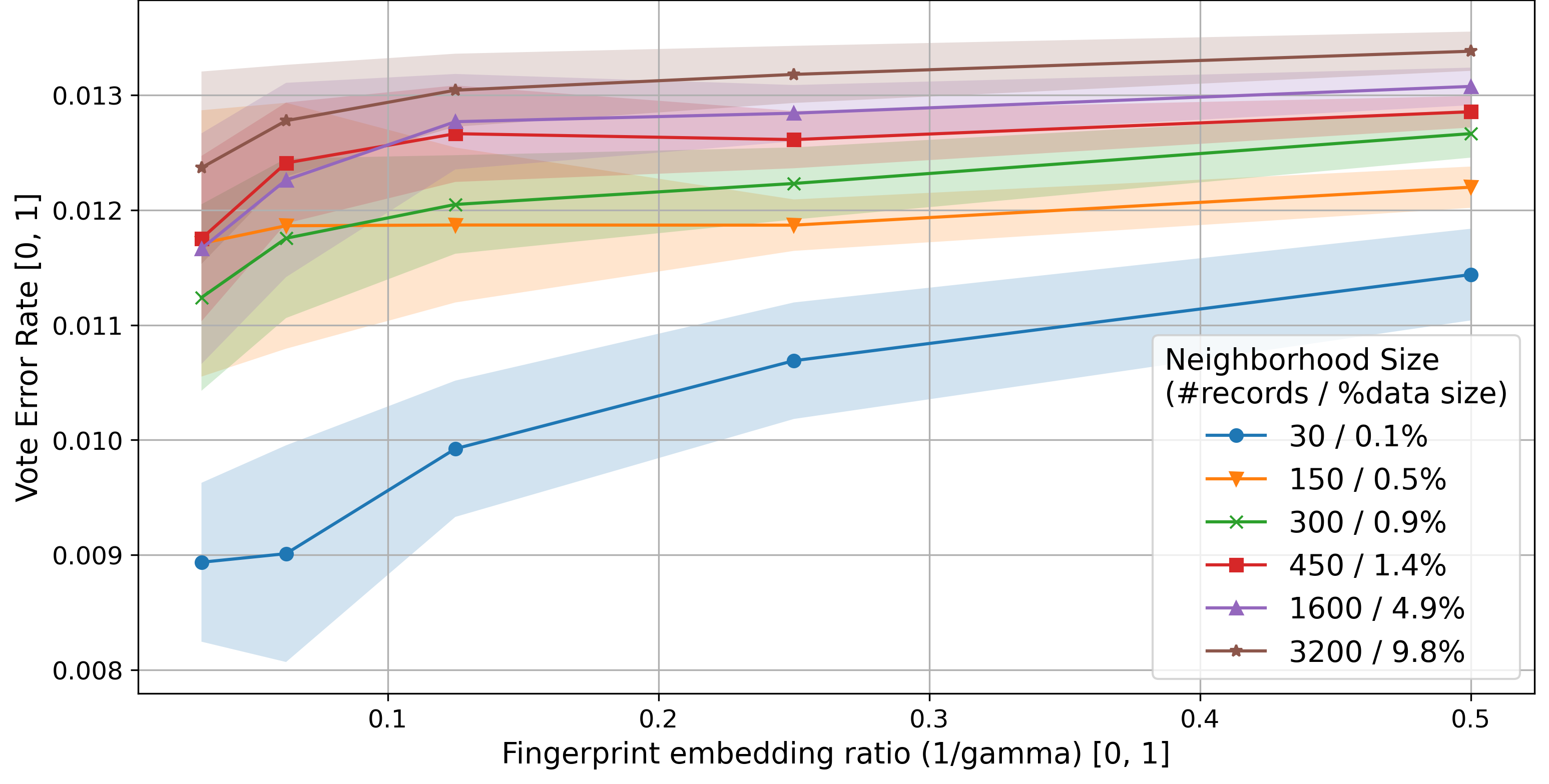}
    \caption{Effectiveness NCorr-FP: Vote Error Rate (VER) on Adult Census dataset for \textit{L}=128 and \textit{N}=20.}
    \label{fig:vote-errors}
\end{figure}

\begin{figure}[t]
    \centering
    \includegraphics[width=\linewidth]{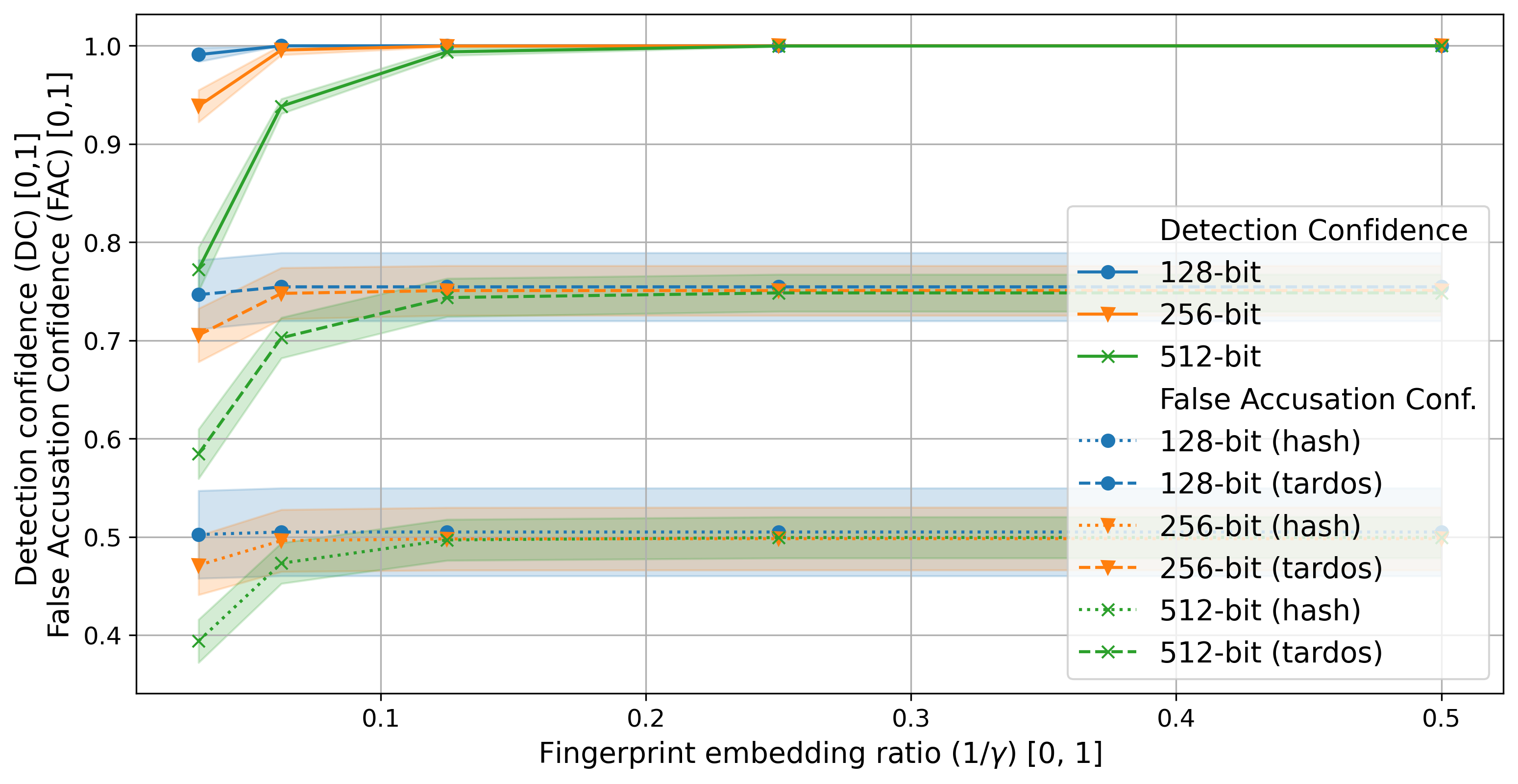}
    \caption{Effectiveness of NCorr-FP : Detection confidence\textit{ DC} (solid lines) and false accusation (\textit{FAC}) for Adult Census data; \textit{k}=300 and \textit{N}=20. \textit{FAC} is shown for two types of fingerprint codes, hash (dotted lines) and Tardos (dashed lines).}
    \label{fig:detection-confidence}
\end{figure}

We empirically evaluate the Vote Error Rate (VER) for different neighbourhood sizes $k$, as minimising VER directly impacts the fingerprint detection performance. 
Although VER generally increases for more marks embedded, it remains $\approx0.013$ in the worst case, as shown in \Cref{fig:vote-errors}. 
VER also increases for larger $k$, indicating stable local distributions (distribution inside the neighbourhood) after the fingerprinting.
The observed low VER does not impact the detectability of the fingerprint, which is at 100\%, as shown in \Cref{fig:detection-confidence} for $k=300$.

In an ideal scenario, the DC of the correct recipient is 1.0 (i.e. perfect bit-wise matching of a detected fingerprint and the fingerprint of the correct recipient), and the \textit{DC} of all other recipients (i.e. false accusation rate, FAC) is around 0.5, in line with random guessing of the bit-sequence. 
In \Cref{fig:detection-confidence} we can see that the $\approx0.50$ false accusation holds for hash fingerprints. 
However, T-codes are designed to have an intentionally larger bit-overlap for detection of colluders. 
Therefore, the false accusation rate for T-codes remains stable at around 0.75. 
This is an important insight for selecting a threshold for the detection confidence when using T-codes.

\subsection{Fidelity}\label{subsec:results-fidelity}
Data accuracy $\mathcal{A}$ is lower-bounded by \Cref{eq:data-accuracy}. 
Still, $\mathcal{A}$ is, in practice, higher due to the sampling procedure we conduct during marking, which allows the value to stay the same. 
This is shown in \Cref{tab:data_accuracy}. 
NCorr-FP also results in higher $\mathcal{A}$ than the baseline fingerprinting.

\begin{table}[ht]
    \centering
    \caption{Fidelity: data accuracy $\mathcal{A}$ of Adult Census dataset. Greater accuracy means more similarity to the original data. The first row is the theoretical lower bound, followed by the baseline random fingerprinting and NCorrFP.}
     \label{tab:data_accuracy}
    \begin{tabular}{l|rrrrr}
    \toprule
        method & $\gamma=32$ & $\gamma=16$ & $\gamma=8$ & $\gamma=4$ & $\gamma=2$ \\
    \midrule
        $1/\nu\gamma$ & 0.9979 & 0.9958 & 0.9917 & 0.9833 & 0.9667 \\
         baseline & 0.9986 & 0.9971 & 0.9942 & 0.9885 & 0.9770 \\
         NCorrFP & 0.9988 & 0.9976 & 0.9952 & 0.9904 & 0.9808 \\
    \end{tabular}
\end{table}

\begin{figure}[ht]
    \centering
    \centering
    \subfloat[\label{fig:histogram-003}Emb. ratio $1/\gamma=0.03$]{%
       \includegraphics[height=125pt]{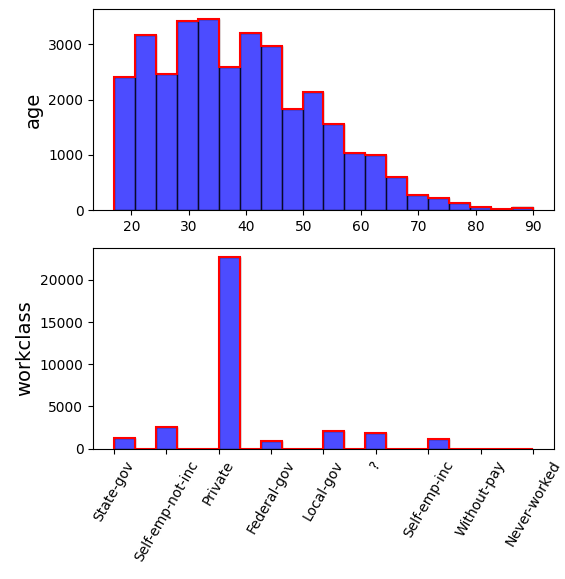}}
    \hfill
        \subfloat[\label{fig:histogram-05}Emb. ratio $1/\gamma=0.5$]{%
           \includegraphics[height=125pt]{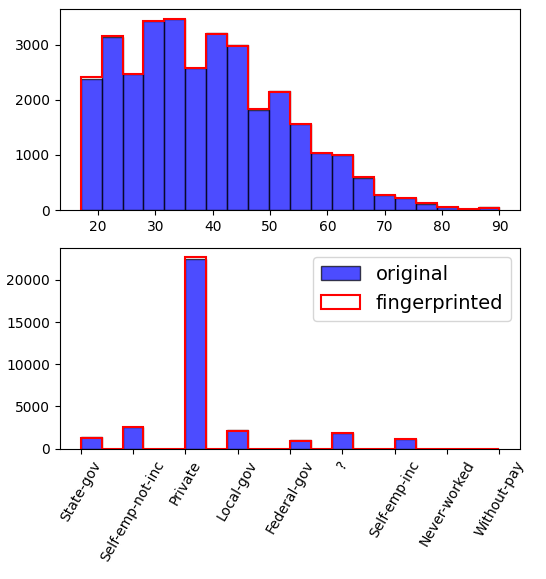}}
    \caption{NCorr-FP fidelity: Histogram difference for 2 representative attributes of the Adult census dataset. 
    \textit{Age} is binned in 20 bins. The original histogram is marked in blue, while the histogram of fingerprinted data is outlined in red.}
    \label{fig:histogram-diff-adult}
\end{figure}

\begin{table}[ht]
    \caption{Fidelity: Hellinger distance between the attributes of the original and corresponding attributes of a fingerprinted dataset. The value range is [0,1] where 0 indicates the identical distributions and 1 completely disjoint distributions.}
    \label{tab:hellinger_adult}
    \resizebox{\linewidth}{!}{
    \centering
    \begin{tabular}{r|r|rrr|rrrr||r}
    1/$\gamma$ & \textit{k} & \textit{age} & \textit{capital-gain} & \textit{capital-loss} & \textit{education} & \textit{m-status} & \textit{race} & \textit{country} & \textit{agg. mean} \\
    \midrule
    0.03 & base & 0.0001 & $5\times10^{-7}$ & $8\times10^{-6}$ & 0.0018 & 0.0044 & 0.0029 & 0.0950 & 0.0286 \\
         & 300 & 0.0005 & 0.0101 & 0.0048 & 0.0003 & 0.0016 & 0.0007 & 0.0013 & 0.0019 \\
         & 450 & 0.0004 & 0.0099 & 0.0051 & 0.0003 & 0.0047 & 0.0007 & 0.0014 & 0.0021 \\
    \midrule
    0.06 & base & 0.0002 & $8\times10^{-7}$ & $2\times10^{-5}$ & 0.0035 & 0.0079 & 0.0055 & 0.0952 & 0.0446 \\
         & 300 & 0.0007 & 0.0238 & 0.0087 & 0.0003 & 0.0042 & 0.0015 & 0.0024 & 0.0037 \\
         & 450 & 0.0008 & 0.0215 & 0.0088 & 0.0004 & 0.0089 & 0.0015 & 0.0024 & 0.0040 \\
    \midrule
    0.13 & base & 0.0003 & $2\times10^{-6}$ & $3\times10^{-5}$ & 0.0066 & 0.0147 & 0.0104 & 0.0962 & 0.0717 \\
         & 300 & 0.0011 & 0.0466 & 0.0181 & 0.0006 & 0.0098 & 0.0029 & 0.0043 & 0.0063 \\
         & 450 & 0.0013 & 0.0459 & 0.0189 & 0.0010 & 0.0152 & 0.0029 & 0.0042 & 0.0070 \\
    \midrule
    0.25 & base & 0.0004 & $3\times10^{-6}$ & $6\times10^{-5}$ & 0.0126 & 0.0258 & 0.0202 & 0.0991 & 0.1126 \\
         & 300 & 0.0015 & 0.0877 & 0.0378 & 0.0007 & 0.0150 & 0.0057 & 0.0085 & 0.0125 \\
         & 450 & 0.0017 & 0.0857 & 0.0390 & 0.0018 & 0.0242 & 0.0057 & 0.0084 & 0.0134 \\
    \midrule
    0.50 & base & 0.0006 & $6\times10^{-6}$ & 0.0001 & 0.0234 & 0.0441 & 0.0377 & 0.1071 & 0.1968 \\
         & 300 & 0.0020 & 0.1412 & 0.0731 & 0.0010 & 0.0243 & 0.0122 & 0.0170 & 0.0211 \\
         & 450 & 0.0022 & 0.1363 & 0.0757 & 0.0030 & 0.0402 & 0.0123 & 0.0167 & 0.0225 \\
    \end{tabular}
    }
\end{table} 
The histograms of data attributes stay well-preserved as we demonstrate in \Cref{fig:histogram-diff-adult}, even towards large embedding ratios of 0.5. 
To quantify these albeit minor distribution shifts, the Hellinger distance \Cref{tab:hellinger_adult} and KL divergence \Cref{tab:kl_divergence_adult} are obtained per attribute and aggregated for the entire dataset.
We can observe very small values for the Hellinger distance, suggesting minimal shifts in attribute distributions. As expected, smaller distribution shifts are observed for smaller embedding ratios and smaller \textit{k}. 
Compared to the baseline fingerprint based on randomised embedding, numerical features consistently show a larger distribution shift for NCorr-FP. This is an expected behaviour by design: the random technique modifies the values by LSB flipping, which results in minor absolute value changes, whereas our similarity-based sampling might result in new values significantly different from the original -- but fitting the context better. 
The distribution of categorical values is, on the other hand, preserved much better via NCorr-FP. 
The same trend follows for KL divergence, where increasing \textit{k} has a slightly lower impact. 

\begin{table}[ht]
    \caption{Fidelity: KL divergence of fingerprinted data attributes from the original reference data attributes. Value range is [0, $\infty$) where 0 indicates identical distributions.}
    \label{tab:kl_divergence_adult}
\resizebox{\linewidth}{!}{
    \centering
    \begin{tabular}{r|r|rr|rrrr||r}
    1/$\gamma$ & \textit{k} & \textit{age} & \textit{capital-loss} & \textit{education} & \textit{m-status} & \textit{race} & \textit{country} & agg. mean \\
    \midrule
        0.03 & base & $9.7\times10^{-8}$ & $3.1\times10^{-10}$ & $1.3\times10^{-5}$ & $7.7\times10^{-5}$ & $3.4\times10^{-5}$ & 0.3226 & 0.0894 \\
    & 300 & $8.8\times10^{-7}$ & $9.1\times10^{-5}$ & $3.3\times10^{-7}$ & $1.0\times10^{-5}$ & $1.8\times10^{-6}$ & $6.7\times10^{-6}$ & $2.6\times10^{-5}$ \\
         & 450 & $7.9\times10^{-7}$ & $1.0\times10^{-4}$ & $4.5\times10^{-7}$ & $8.3\times10^{-5}$ & $1.8\times10^{-6}$ & $8.2\times10^{-6}$ & $2.7\times10^{-5}$ \\
    \midrule
        0.06 & base & $1.9\times10^{-7}$ & $1.2\times10^{-9}$ & $5.0\times10^{-5}$ & $2.3\times10^{-4}$ & $1.2\times10^{-4}$ & 0.3228 & 0.1287 \\
    & 300 & $2.1\times10^{-6}$ & $3.0\times10^{-4}$ & $3.3\times10^{-7}$ & $6.5\times10^{-5}$ & $9.0\times10^{-6}$ & $2.2\times10^{-5}$ & $1.3\times10^{-4}$ \\
         & 450 & $2.6\times10^{-6}$ & $3.1\times10^{-4}$ & $7.3\times10^{-7}$ & $2.8\times10^{-4}$ & $9.0\times10^{-6}$ & $2.3\times10^{-5}$ & $1.4\times10^{-4}$ \\
    \midrule
        0.13 & base & $4.2\times10^{-7}$ & $4.2\times10^{-9}$ & $1.7\times10^{-4}$ & $7.5\times10^{-4}$ & $4.3\times10^{-4}$ & 0.3234 & 0.2391 \\
    & 300 & $5.2\times10^{-6}$ & $1.3\times10^{-3}$ & $1.3\times10^{-6}$ & $3.3\times10^{-4}$ & $3.3\times10^{-5}$ & $7.5\times10^{-5}$ & $4.4\times10^{-4}$ \\
         & 450 & $6.6\times10^{-6}$ & $1.4\times10^{-3}$ & $4.1\times10^{-6}$ & $7.7\times10^{-4}$ & $3.3\times10^{-5}$ & $7.1\times10^{-5}$ & $4.8\times10^{-4}$ \\
    \midrule
        0.25 & base & $8.2\times10^{-7}$ & $1.6\times10^{-8}$ & $6.0\times10^{-4}$ & $2.2\times10^{-3}$ & $1.5\times10^{-3}$ & 0.3254 & 0.4654 \\
   & 300 & $8.5\times10^{-6}$ & $5.5\times10^{-3}$ & $2.0\times10^{-6}$ & $7.5\times10^{-4}$ & $1.3\times10^{-4}$ & $2.9\times10^{-4}$ & $2.0\times10^{-3}$ \\
         & 450 & $1.1\times10^{-5}$ & $5.8\times10^{-3}$ & $1.4\times10^{-5}$ & $1.8\times10^{-3}$ & $1.3\times10^{-4}$ & $2.9\times10^{-4}$ & $2.1\times10^{-3}$ \\
    \midrule
        0.50 & base & $1.6\times10^{-6}$ & $6.1\times10^{-8}$ & $2.0\times10^{-3}$ & $6.0\times10^{-3}$ & $5.2\times10^{-3}$ & 0.3310 & 1.0058 \\
    & 300 & $1.6\times10^{-5}$ & $1.9\times10^{-2}$ & $3.8\times10^{-6}$ & $1.8\times10^{-3}$ & $5.9\times10^{-4}$ & $1.1\times10^{-3}$ & $5.7\times10^{-3}$ \\
         & 450 & $2.0\times10^{-5}$ & $2.1\times10^{-2}$ & $3.6\times10^{-5}$ & $4.7\times10^{-3}$ & $5.9\times10^{-4}$ & $1.1\times10^{-3}$ & $6.0\times10^{-3}$ \\
    \end{tabular}
    }
\end{table} 

\begin{figure*}
    \centering
    \includegraphics[width=\linewidth]{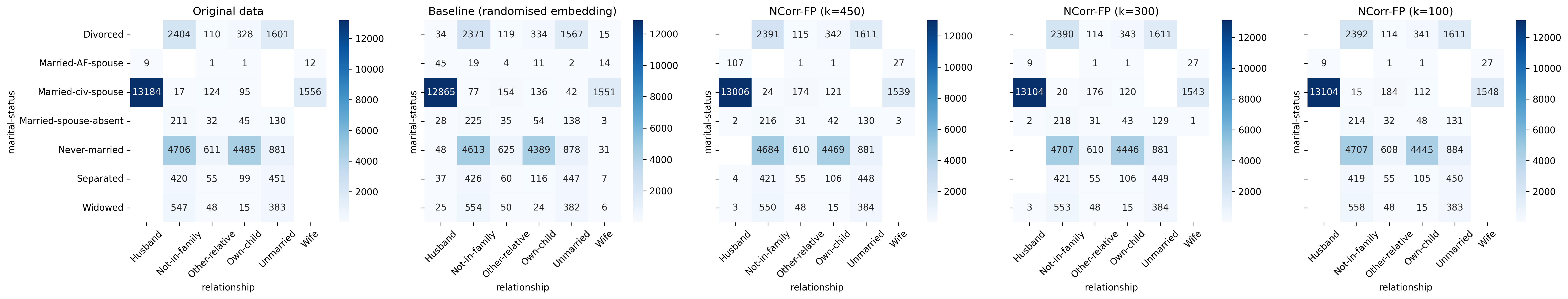}
    \caption{Fidelity: Pairwise histogram for attributes \textit{marital-status}/\textit{relationship}. NCorr-FP preserves the histograms of the original data better than the random embedding. Additionally, adjusting (decreasing) the neighbourhood size \textit{k} of NCorr-FP diminishes the occurrence of the unlikely value combinations, for a fixed embedding ratio $1/\gamma=0.25$.}
    \label{fig:pairwise-hist-msr}
\end{figure*}

We further specifically analyse the variables' interactions to measure the effects on fidelity that do not occur in univariate statistics.
One of the core goals of NCorr-FP is avoiding uncommon value combinations, which is particularly useful when dealing with highly correlated (categorical) data, though it can frequently occur for numerical data as well. 
We can observe this from the pairwise histograms in the Adult Census dataset, for instance, \textit{marital-status}/\textit{relationship} in \Cref{fig:pairwise-hist-msr} 

\begin{figure*}
    \centering
    \includegraphics[width=\linewidth]{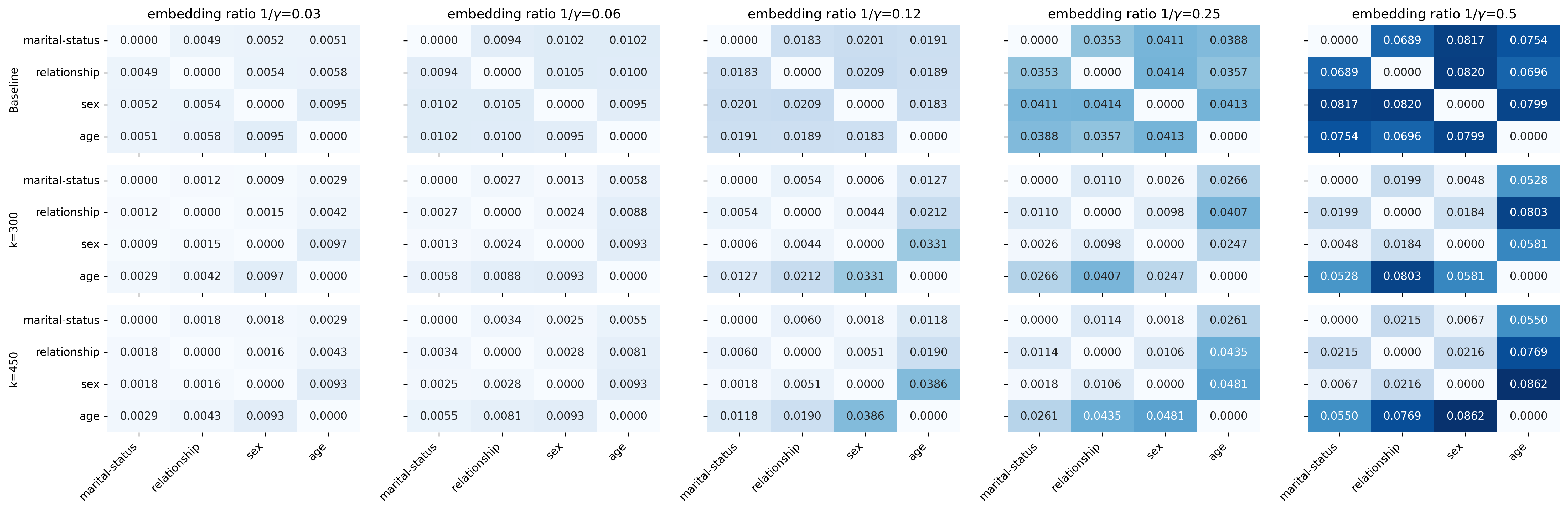}
    \caption{Attribute correlation change$^3$ 
    between the original and fingerprinted Adult Census data. The subset of attributes is used for brevity. The top row represents the baseline fingerprinting and the other NCorr scheme with \textit{k}=300 and \textit{k}=450, respectively. Darker shades represent more change in the attribute correlations. }
    \label{fig:correlation-heatmaps}
\end{figure*}

We also look at the correlations of the attributes\footnote{In the correlation maps of attributes, we use the metric appropriate for the attribute types involved, either Pearson's coefficient, Cramer's V or $\eta^2$. See: \Cref{subsec:evaluation-methodology-fidelity}: \textit{Multivariate and correlation statistics}.\label{fn:corr}}. 
\Cref{fig:correlation-heatmaps} demonstrates the improvement in preserving the correlations by using NCorr embedding (two bottom rows) compared the the baseline (top row).
Binary attributes such as \textit{sex} represent a challenge in preserving the correlations due to their limited domains. This is why, in some cases, the correlation with these attributes is more disrupted than the others - an example is the \textit{age}/\textit{sex} correlation. 
Nevertheless, even the highest correlation shifts are $<10\%$ (note that this is the \textit{relative} correlation change).

\subsection{Utility}\label{subsec:results-utility}
\begin{figure}
    \centering
    \includegraphics[width=0.95\linewidth]{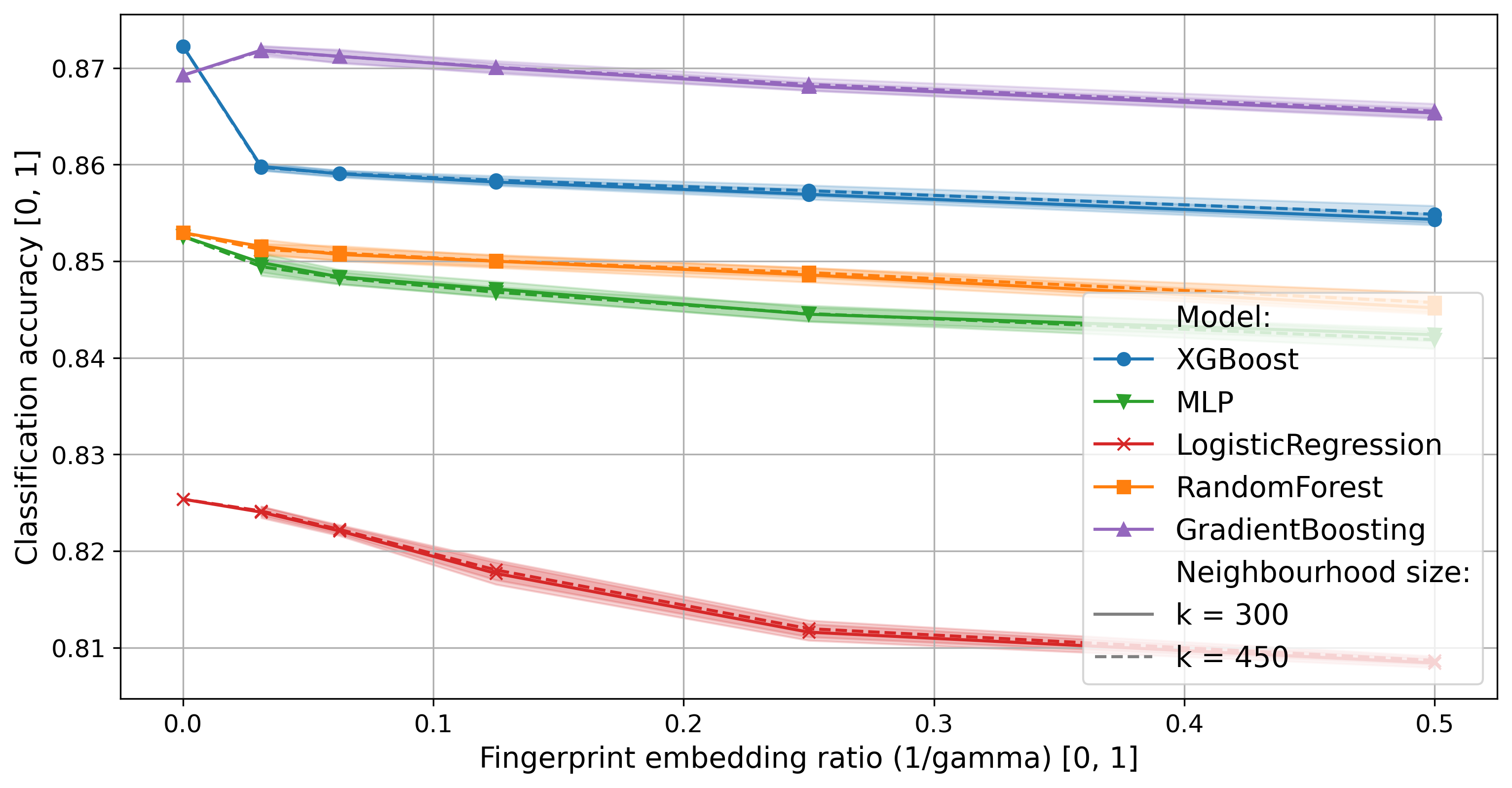}
    \caption{Utility of the fingerprinted data: ML classification accuracy via 4 classifiers. An embedding ratio of 0.0 represents the original dataset.}
    \label{fig:utility}
\end{figure}
\Cref{fig:utility} shows that introducing fingerprints leads to a gradual degradation in predictive performance. However, the drop in accuracy remains relatively small, indicating that the utility of the data is largely preserved. 
Among the four classifiers, XGBoost consistently achieved the highest accuracy and decreased in performance at most by 0.016 (1.6\%). Logistic Regression shows the same magnitude of reduction, however the absolute accuracy score is significantly lower compared to XGBoost.
The greatest robustness to fingerprinting was demonstrated by Random Forest, for which the classification accuracy decreased at most 0.008 (0.8\%).  
MLP also showed strong resilience, with accuracy reductions around 1\%, comparable to Random Forest.
Interestingly, the results are largely consistent across the two tested neighborhood sizes (k = 300 and k = 450), suggesting that the embedding strength, rather than the neighborhood size, has a more pronounced effect on utility.

Overall, the fingerprinting method maintains high data utility for most practical use cases, exhibiting stability in experiments, evident from low standard deviations in classification accuracy. 

\subsection{Robustness}\label{subsec:results-robustness}
\subsubsection{Single-user attacks}\label{subsubsec:results-robustness-single-user}

\begin{figure}
    \centering
    \subfloat[\label{fig:robustness-subset-horizontal}Horizontal]{%
       \includegraphics[height=94pt]{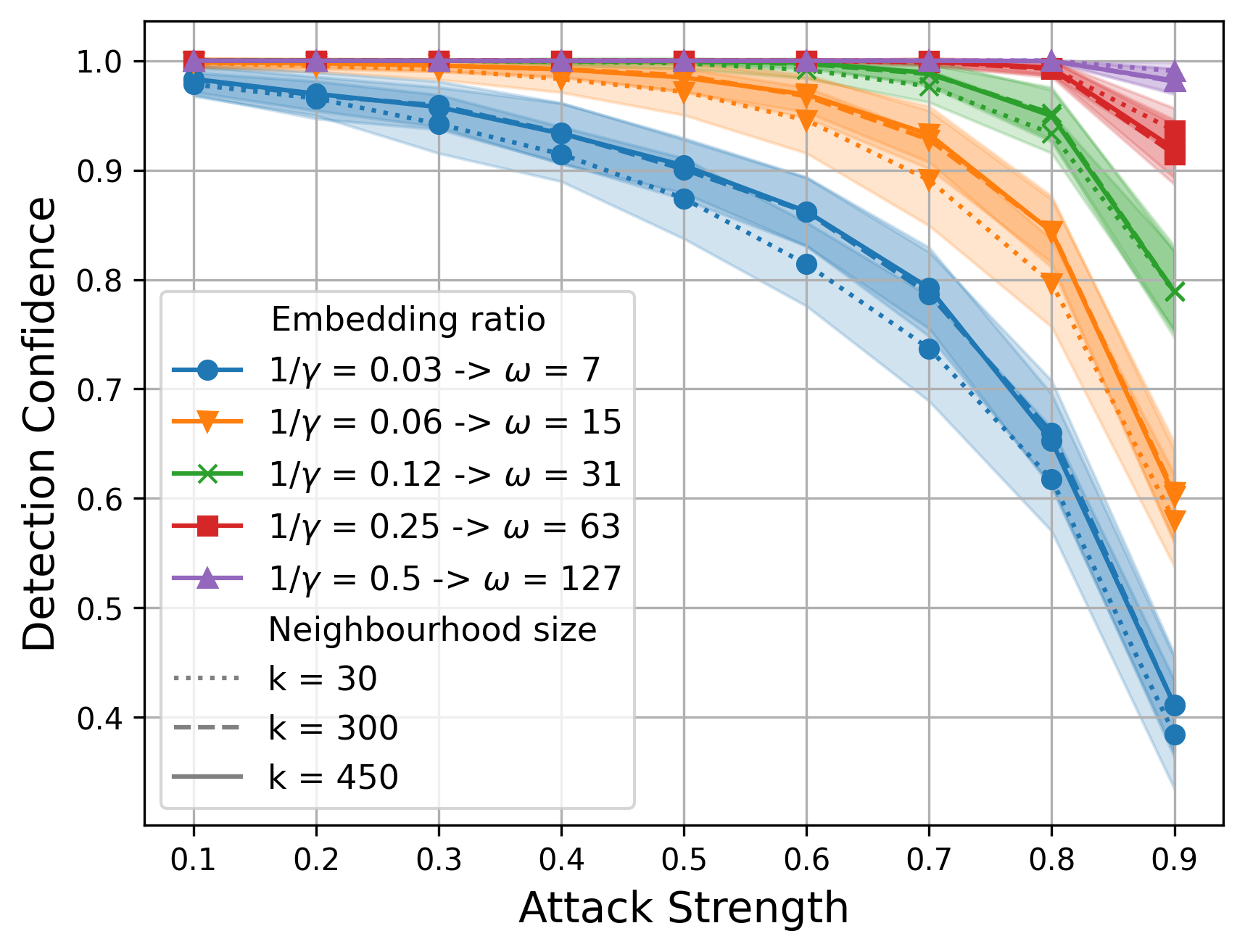}}
    \hfill
    \subfloat[\label{fig:robustness-subset-vertical}Vertical]{
       \includegraphics[height=94pt]{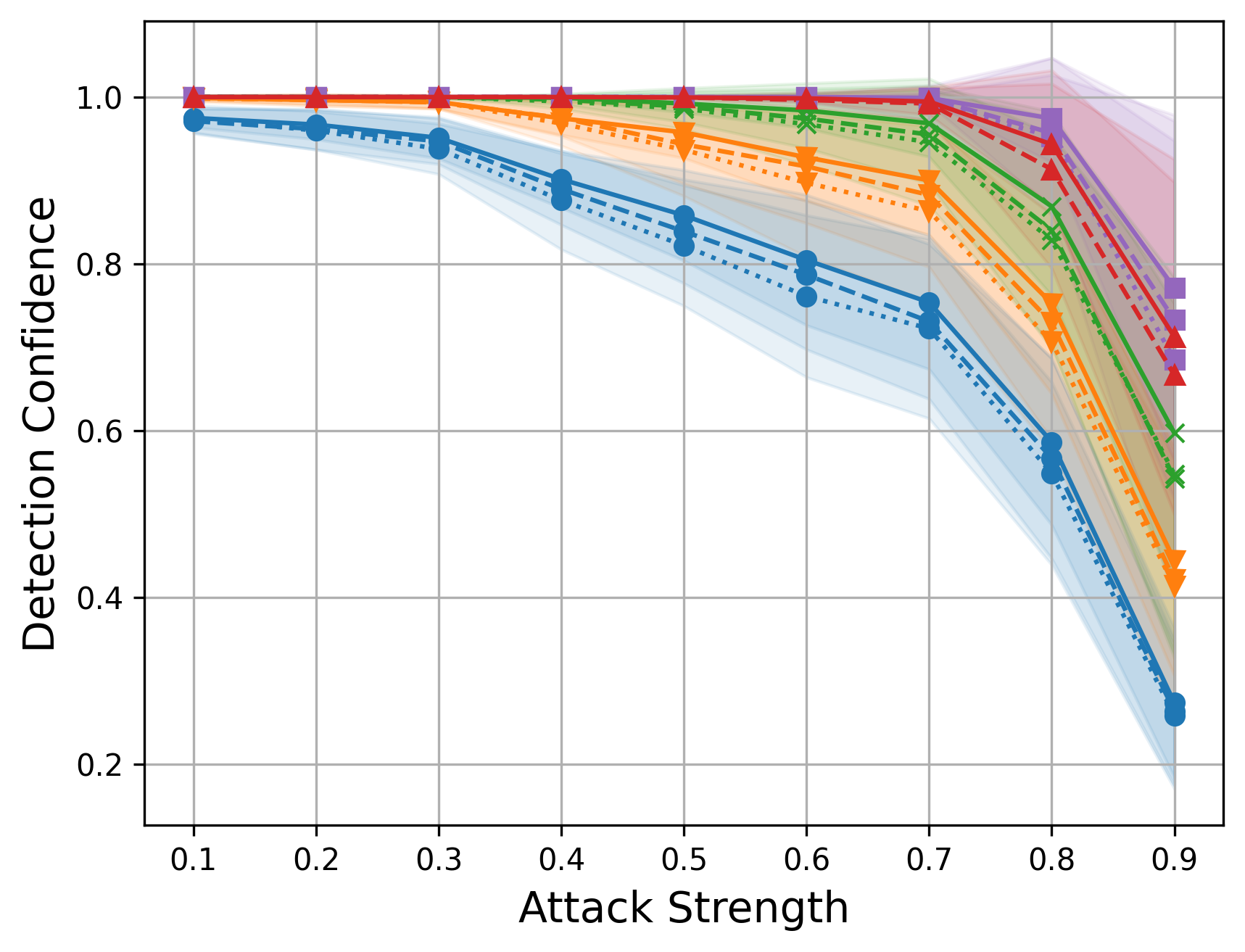}}
    \caption{Robustness of NCorr-FP: \textit{DC} under subsetting attacks}
    \label{fig:robustness-subset}
\end{figure}

Horizontal (\Cref{fig:robustness-subset-horizontal}) and vertical subsetting (\Cref{fig:robustness-subset-vertical}) can disrupt the confidence of fingerprint extraction depending on the fingerprint parameter choice. 
While the neighbourhood size \textit{k} shows little influence on robustness against these attacks, the embedding ratio $1/\gamma$ plays a more significant role; increasing it leads to more robust fingerprints. 
Taking into account the length of the fingerprints used in these experiments, $L=128$, we can obtain the fingerprint-bit redundancies $\omega$ that are essential to robustness. 
For instance, with $\omega=127$, the fingerprint is extracted with perfect confidence even from 20\% of remaining data samples and 30\% remaining data columns. 

\Cref{tab:attack_loss} summarises the cost of the horizontal attack according to the fidelity measures Hellinger distance and KL divergence in worst-case scenarios for the defender: lower range of embedding ratio (weaker fingerprints) and upper range of attack strength (stronger attacks).
Empirical results show that all of the most disruptive attacks (those that decrease the extraction confidence below 95\%), marked in bold, are also reducing fidelity more than it has been reduced due to the fingerprint embedding (c.f. last column Fidelity $\mathcal{R}'$). 
In the case of KL divergence, the difference in fidelity is a few orders of magnitude. 
Hence, the attacker bears the cost of additional fidelity loss, which might serve as a deterrent for performing more severe attacks. 
This attacker's cost further reinforces the robustness of the scheme, as undermining the fingerprint comes at a higher fidelity loss than embedding it.

\begin{table}[t]
    \centering
    \caption{Horizontal attack cost: fidelity loss for the attacker in horizontal subsetting. We show $k=300$, less robust scenarios with $1/\gamma \in [0.03, 0.06, 0.13]$ and stronger attacks of $>50\%$ for brevity. Attacks that reduce \textit{DC} to $<95\%$ are bolded.
    Fidelity of an unchanged fingerprinted dataset $\mathcal{R}'$ is a baseline for comparison.}
    \label{tab:attack_loss}
    \resizebox{\linewidth}{!}{
        \begin{tabular}{l|lrrrrr|r}
        \toprule
        $1/\gamma$ & \textit{attack\_strength}: & 0.5 & 0.6 & 0.7 & 0.8 & 0.9 & fidelity $\mathcal{R}'$\\
        \midrule
        0.03 & Hellinger dist. & \textbf{0.0183} & \textbf{0.0234} & \textbf{0.0296 }& \textbf{0.0379} & \textbf{0.0528} & 0.0019 \\
        &  KL divergence & \textbf{0.0031} &	\textbf{0.0051} & \textbf{0.0081} & \textbf{0.0131 }&	\textbf{0.0252} & $2.6\times10^{-5}$ \\ 
        \midrule
        0.06 & Hellinger dist. & 0.0194	&0.0244&	\textbf{0.030}6 &	\textbf{0.0389}&	\textbf{0.0538} & 0.0037 \\
         &  KL divergence & 0.0036 &	0.0057  &	\textbf{0.0088} &	\textbf{0.0139}	& \textbf{0.0262} & $1.3\times10^{-4}$ \\
       \midrule
       0.13 & Hellinger dist. & 0.0216 & 0.0265 & 0.0326 &	0.0407 & \textbf{0.0555} & 0.0063 \\
       & KL divergence & 0.0047 &	0.0070&	0.0103	& 0.0156	& \textbf{0.0283} & $4.4\times10^{-4}$ \\
        \bottomrule
        \end{tabular}
    }
\end{table}  

\begin{figure*}
    \centering
    \subfloat[\label{fig:robustness-flipping-random}Random flipping]{%
       \includegraphics[height=110pt]{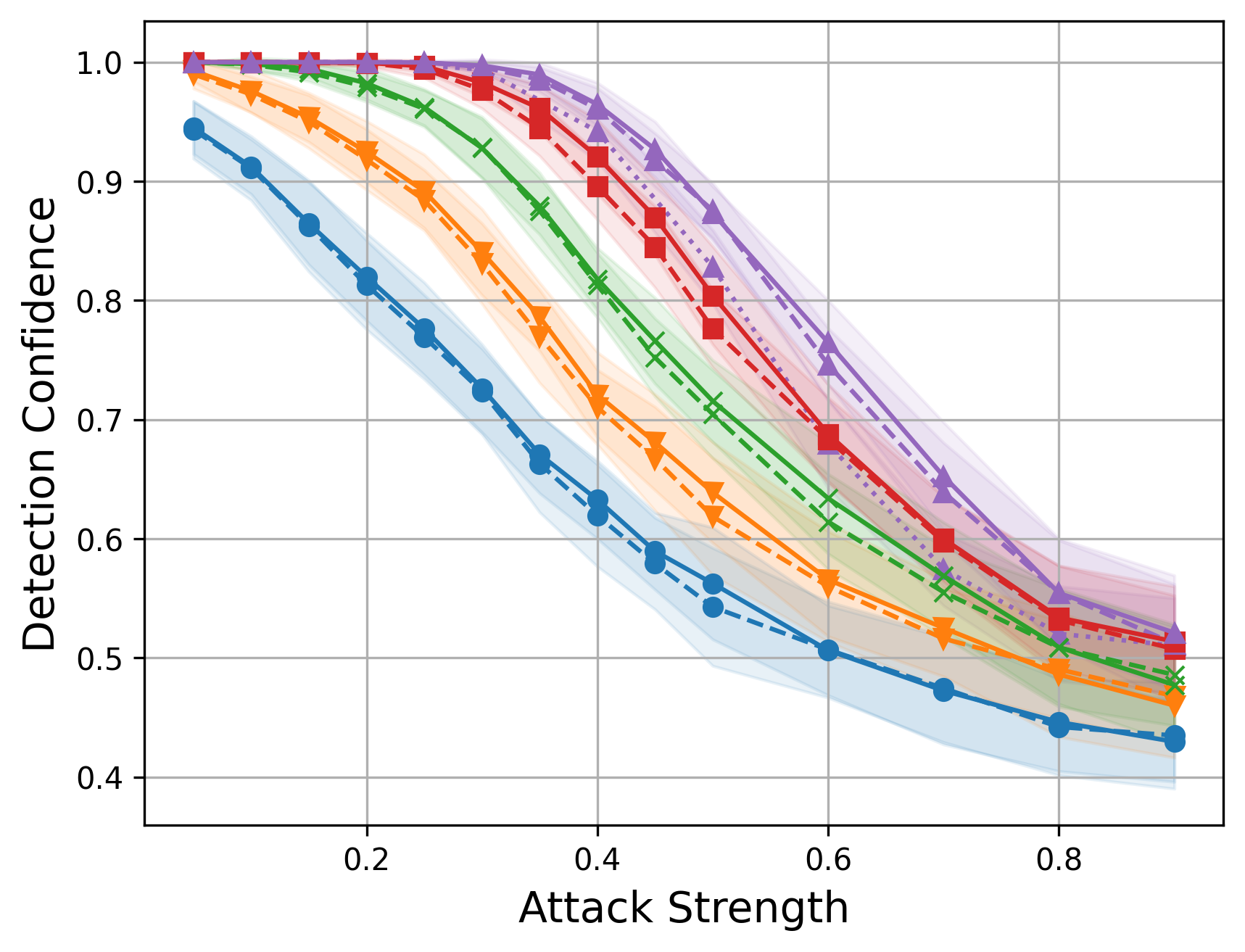}}
    \hfill\subfloat[\label{fig:robustness-flipping-cluster}Cluster flipping]{%
       \includegraphics[height=110pt]{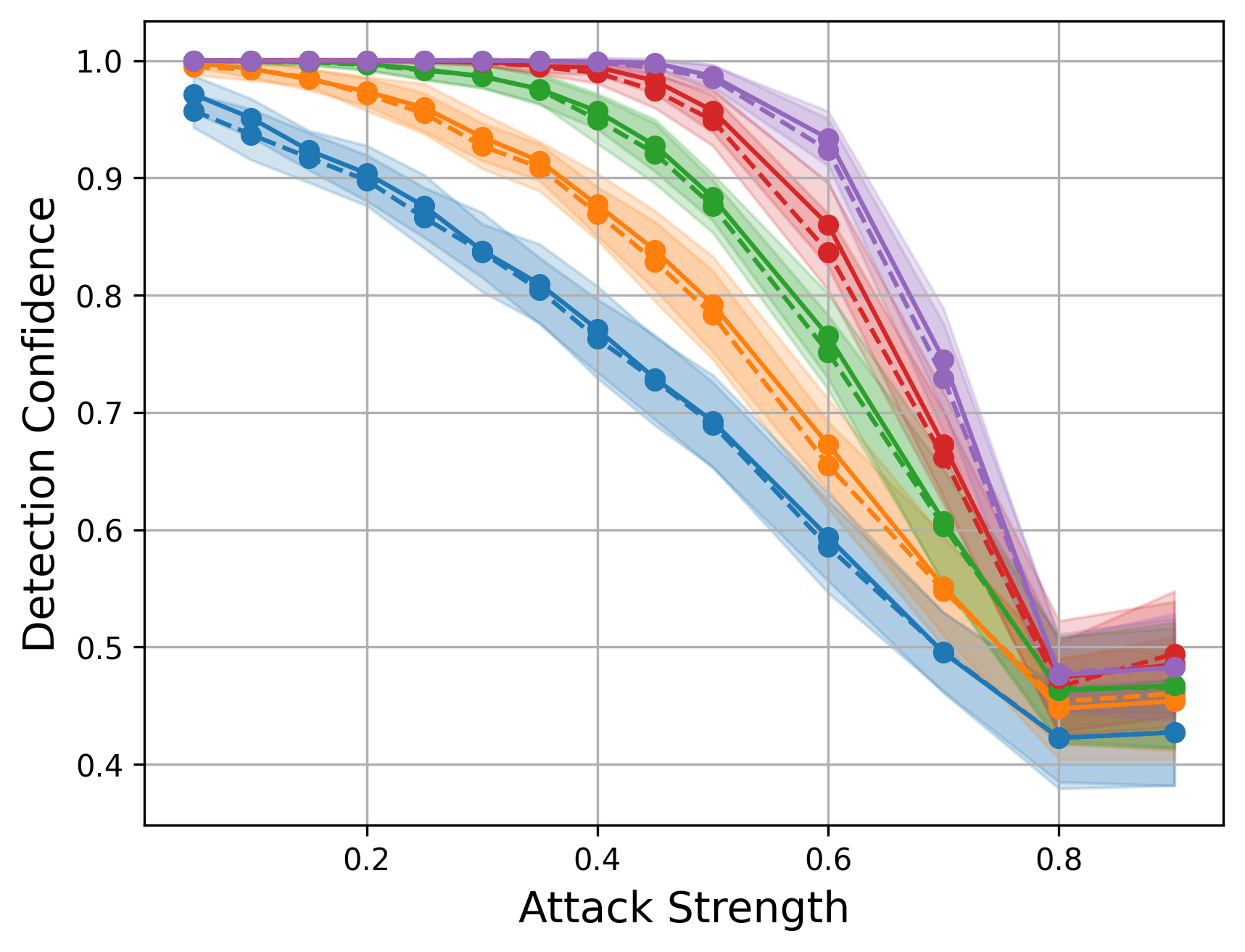}}
    \hfill
    \subfloat[\label{fig:robustness-flipping-cluster-ep}Cluster flipping with exact params]{%
       \includegraphics[height=110pt]{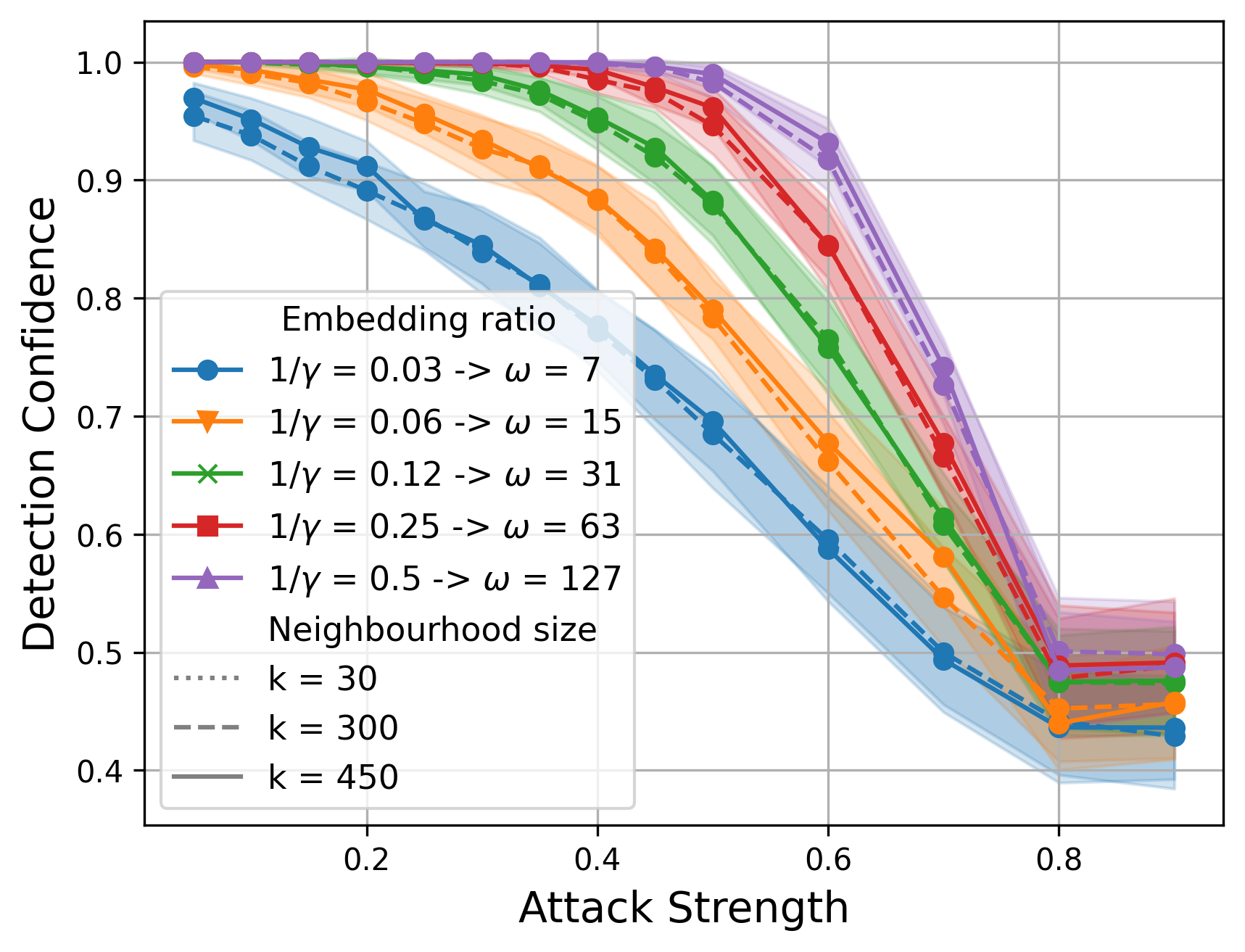}}
    \caption{Robustness of NCorr-FP: Detection Confidence (DC) under flipping attacks.}
    \label{fig:robustness-flipping}
\end{figure*}

\begin{table}[t]
    \centering
    \caption{Flipping attack cost: fidelity loss for the attacker in flipping attack. We show $k=300$, less robust scenarios with $1/\gamma\in[0.03,0.06,0.13]$ and attacks of strength up to 50\% for brevity. Attacks that reduce \textit{DC} to $<95\%$ are bolded.
    Fidelity of an unchanged fingerprinted dataset $\mathcal{R}'$ is a baseline for comparison.}
    \label{tab:attack_loss_f}
    \resizebox{\linewidth}{!}{
        \begin{tabular}{l|lrrrrr|r}
        \toprule
        $1/\gamma$ & \textit{attack\_strength}: & 0.1 & 0.2 & 0.3 & 0.4& 0.5& fidelity $\mathcal{R}'$\\
        \midrule
        0.03 & Hellinger dist. & \textbf{0.0849} & \textbf{0.1390} & \textbf{0.1842} & \textbf{0.2249} &\textbf{ 0.2634} & 0.0019 \\
             & KL divergence   & \textbf{0.0370} & \textbf{0.0893} &\textbf{ 0.1491} & \textbf{0.2173} &\textbf{ 0.2956} & $2.6\times10^{-5}$ \\
        \midrule
        0.06 & Hellinger dist. & 0.0854 & \textbf{0.1394} &\textbf{ 0.1845} &\textbf{ 0.2252} & \textbf{0.2635} & 0.0037 \\
             & KL divergence   & 0.0375 & \textbf{0.0899} & \textbf{0.1498} & \textbf{0.2178} &\textbf{ 0.2960} & $1.3\times10^{-4}$ \\
       \midrule
        0.13 & Hellinger dist. & 0.0867 & 0.1403 & \textbf{0.1851} & \textbf{0.2256} & \textbf{0.2639} & 0.0063 \\
        & KL divergence &  0.0388 &  0.0912&  \textbf{0.1510} &  \textbf{0.2190} &  \textbf{0.2970} & $4.4\times10^{-4}$ \\
        \bottomrule
        \end{tabular}
    }
\end{table}

The results for three types of flipping attacks are compared in \Cref{fig:robustness-flipping}: (i) randomised value flipping, and flipping of values only within the subset (cluster) of the most influential records, selected (ii) with approximated or (iii) exact fingerprint parameter setting.
Flipping data values reduces the confidence of fingerprint detection at a higher rate compared to the subsetting attacks; however, increasing the embedding ratio leads to relatively robust settings. 
For instance, with an embedding ratio $1/\gamma=0.13$, the fingerprint has a $>95\%$ detection confidence (DC) for up to 25\% of flipped data values, while for a higher $1/\gamma=0.5$ the robustness with $DC>95\%$ is preserved for up to 40\% of flipped values. 
The flipping attack cost is shown in \Cref{tab:attack_loss_f}. 
For the particular example above with $1/\gamma=0.13$ and attack strength of $>25\%$ (i.e. the attack that successfully reduces the detection confidence to below 95\% -- in \Cref{tab:attack_loss_f} observing the $attack\_strength=0.3$), the fidelity cost is two (Hellinger) and three (KL divergence) orders of magnitude higher than that of embedding the fingerprint in the data.
This general trend follows for all other settings in \Cref{tab:attack_loss_f}. 
Particularly relevant are the attack settings (including the example) where the attacker reduces DC to some degree, e.g. 95\%, however also obtains a high fidelity loss (those are bolded in \Cref{tab:attack_loss_f}). 
Hence, a successful flipping attack always comes at a high fidelity cost.

The cluster flipping attack (\Cref{fig:robustness-flipping-cluster,fig:robustness-flipping-cluster-ep}) is outperformed by the randomised value flipping attack (\Cref{fig:robustness-flipping-random}).
Although the malicious modifications are directed towards the most influential records in the neighbourhood calculation during the detection process, the distributions in these neighbourhoods still are sufficiently stable not to get disrupted by these changes. 
By accumulating the modifications towards specific (and overall fewer) records, the adaptive attack loses the opportunity to modify and likely remove fingerprint bits from other records. Hence, distributing the malicious modifications randomly and uniformly across the dataset proves to be a more successful strategy for removing the fingerprint. 

\subsubsection{Collusion attacks}\label{subsubsec:results-robustness-collusion}
\begin{table*}[t]
    \centering
    \tiny
    \caption{Collusion attacks via three strategies: (i) averaging, (ii) substitution and (iii) substitution + flip. We use \textit{N}=20 and Tardos accusation mechanism with $Z_1$. Bolded are the settings with perfect $precision=1.0$.}
    \label{tab:robustness-collusion}
    \resizebox{\linewidth}{!}{
    \begin{tabular}{rr|lll|lll|lll}
    \toprule
    \textit{c} & \textit{L} 
    & \multicolumn{3}{c|}{Average} 
    & \multicolumn{3}{c|}{Substitution} 
    & \multicolumn{3}{c}{Substitution+Flip} \\
    & 
    & precision $\uparrow$ & FAR $\downarrow$ & recall $\uparrow$
    & precision $\uparrow$ & FAR $\downarrow$ & recall $\uparrow$
    & precision $\uparrow$ & FAR $\downarrow$ & recall $\uparrow$ \\
    \midrule
    2 & 128  & \textbf{1.00±0.00} & \textbf{0.00±0.00} & \textbf{1.00±0.00}  & 0.90±0.16 & 0.10±0.16 & 1.00±0.00 & 0.97±0.11 & 0.03±0.11 & 1.00±0.00 \\
    2 & 256  & \textbf{1.00±0.00} & \textbf{0.00±0.00} & \textbf{1.00±0.00}  & \textbf{1.00±0.00} & \textbf{0.00±0.00} & \textbf{1.00±0.00} & \textbf{1.00±0.00} & \textbf{0.00±0.00} & \textbf{1.00±0.00} \\
    2 & 512  & \textbf{1.00±0.00} & \textbf{0.00±0.00} & \textbf{1.00±0.00}  & \textbf{1.00±0.00} & \textbf{0.00±0.00} & \textbf{1.00±0.00} & \textbf{1.00±0.00} & \textbf{0.00±0.00} & \textbf{1.00±0.00} \\
    2 & 1024 & \textbf{1.00±0.00} & \textbf{0.00±0.00} & \textbf{1.00±0.00}  & \textbf{1.00±0.00} & \textbf{0.00±0.00} & \textbf{1.00±0.00} & \textbf{1.00±0.00} & \textbf{0.00±0.00} & \textbf{1.00±0.00} \\
    \addlinespace[2pt]
    3 & 128  & 0.89±0.14 & 0.11±0.14 & 0.87±0.17 & 0.83±0.22 & 0.17±0.22 & 0.83±0.18 & 0.85±0.21 & 0.15±0.21 & 0.90±0.16 \\
    3 & 256  & 0.98±0.08 & 0.03±0.08 & 1.00±0.00 & \textbf{1.00±0.00} & \textbf{0.00±0.00} & \textbf{1.00±0.00} & 0.95±0.11 & 0.05±0.11 & 1.00±0.00 \\
    3 & 512  & \textbf{1.00±0.00} & \textbf{0.00±0.00} & \textbf{1.00±0.00} & \textbf{1.00±0.00} & \textbf{0.00±0.00} & \textbf{1.00±0.00} & \textbf{1.00±0.00} & \textbf{0.00±0.00} & \textbf{1.00±0.00} \\
    3 & 1024 & \textbf{1.00±0.00} & \textbf{0.00±0.00} & \textbf{1.00±0.00} & \textbf{1.00±0.00} & \textbf{0.00±0.00} & \textbf{1.00±0.00} & \textbf{1.00±0.00} & \textbf{0.00±0.00} & \textbf{1.00±0.00} \\
    \addlinespace[2pt]
    5 & 128  & 0.79±0.26 & 0.21±0.26 & 0.48±0.27 & 0.90±0.16 & 0.10±0.16 & 0.62±0.20 & 0.82±0.17 & 0.18±0.17 & 0.58±0.18 \\
    5 & 256  & 0.96±0.10 & 0.04±0.10 & 0.78±0.18 & 0.83±0.19 & 0.18±0.19 & 0.58±0.20 & 0.93±0.11 & 0.07±0.11 & 0.68±0.17 \\
    5 & 512  & 0.98±0.05 & 0.02±0.05 & 0.82±0.15 & \textbf{1.00±0.00} & \textbf{0.00±0.00} & \textbf{0.86±0.13} & 0.98±0.06 & 0.02±0.06 & 0.76±0.13 \\
    5 & 1024 & \textbf{1.00±0.00} & \textbf{0.00±0.00} & 0.82±0.18 & 0.93±0.11 & 0.07±0.11 & 0.76±0.13 & 0.94±0.10 & 0.06±0.10 & 0.72±0.14 \\
    \addlinespace[2pt]
    10 & 128  & 0.82±0.24 & 0.18±0.24 & 0.29±0.12 & 0.73±0.33 & 0.28±0.33 & 0.23±0.12 & 0.71±0.24 & 0.30±0.24 & 0.22±0.11 \\
    10 & 256  & 0.88±0.16 & 0.12±0.16 & 0.33±0.07 & 0.72±0.35 & 0.28±0.35 & 0.27±0.13 & 0.80±0.25 & 0.20±0.25 & 0.26±0.07 \\
    10 & 512  & 0.95±0.11 & 0.05±0.11 & 0.30±0.08 & 0.88±0.16 & 0.13±0.16 & 0.28±0.06 & 0.90±0.16 & 0.10±0.16 & 0.25±0.05 \\
    10 & 1024 & 0.98±0.08 & 0.03±0.08 & 0.32±0.08 & 0.89±0.17 & 0.11±0.17 & 0.27±0.08 & 0.87±0.17 & 0.13±0.17 & 0.31±0.07 \\
    \bottomrule
    \end{tabular}
}
\end{table*}

\Cref{tab:robustness-collusion} shows the resulting analysis on collusion resolution using Tardos codes, cf. \Cref{sec:system}, with accusation threshold $Z_1$ (\Cref{eq:tardos-accusation-threshold}).
It can be observed that the length of the fingerprint is a crucial requirement for increasing confidence in detecting larger collusions. 
Shorter 128-bit fingerprints might be sufficient for settings with a lower number of recipients (hence lower potential colluders). 
For larger collusions, 1024-bit fingerprints still give high precision rates for all attacks, hence there is a high confidence that at least one colluder will be detected. 
It is relevant to note that in our experiments, the highest-scoring accusation was always a true colluder (indicated by a precision score of 1.0). 
However, we can observe significantly lower recall scores for $c>2$, indicating that not all colluders are successfully identified. 
Our empirical results across various collusion strategies show that a 256-bit fingerprint is sufficient to reliably identify two colluders. 
However, as the number of colluders increases, the required fingerprint length grows rapidly, with even 1024-bit fingerprints showing detection errors when attempting to identify just five colluders. 
Despite this, T-codes remain a viable solution for collusion detection, offering reasonable precision under some uncertainty.

\subsection{Summary \& Guidelines}\label{subsec:results-summary}
We summarize the influence of core fingerprinting parameters on the four system requirements in \Cref{tab:summary}. 
The trends are based on empirical observations across multiple experiments. 
Arrows indicate whether a parameter should be increased ($\uparrow$) or decreased ($\downarrow$) to improve a given requirement; “–” denotes no significant observed effect.

\begin{table}[ht]
    \centering
    \caption{Effect of fingerprinting parameters on system requirements.}
    \resizebox{\linewidth}{!}{
    \begin{tabular}{l|cccc}
    \toprule
         param  & EFFECTIVENESS  & FIDELITY & UTILITY  & ROBUSTNESS \\
         \midrule
         \textit{k} & $\downarrow$   & $\downarrow$  & -            & $\uparrow$ \\
         $1/\gamma$ & $\uparrow$     & $\downarrow$  & $\downarrow$ & $\uparrow$ \\
         $L$        & $\downarrow_{DC}\quad\uparrow_{FAC}$ & - & - & $\downarrow_{S}\quad\uparrow_{C}$ \\
         $\omega$ & $\uparrow$ & - & - & $\uparrow$ \\ 
         \bottomrule
    \end{tabular}
    }
    \label{tab:summary}
\end{table}

The redundancy factor $\omega$ captures the relationship between fingerprint length $L$ and embedding ratio $1/\gamma$, and is the main determinant of detection performance.
The results on effectiveness in \Cref{fig:detection-confidence} show that fingerprints with greater length require a higher embedding ratio $1/\gamma$ to achieve perfect detection confidence $DC=100\%$. 
This indicates that neither $L$ nor $1/\gamma$ alone governs effectiveness; rather, their combination, as captured by $\omega=n/L\gamma$, determines whether the fingerprint can be reliably detected.
Empirically, we find that a minimum redundancy of $\omega \geq 16$ is required to consistently achieve $DC = 100\%$. 
While this is a practical lower bound for effectiveness, higher redundancy values are recommended to ensure robustness, particularly under adversarial conditions, as indicated in \Cref{tab:summary}. To increase $\omega$, one can either decrease $L$ or increase $1/\gamma$.

The parameter $L$ has a dual effect: reducing $L$ decreases redundancy $\omega$, which improves \textit{DC} and robustness against single-user attacks ($\downarrow_{S}$), while increasing $L$ reduces false accusation confidence (FAC) and improves collusion resilience ($\uparrow_{C}$). 
Therefore, the design goal is to select the largest possible $L$ such that the redundancy constraint $\omega\geq16$ is still satisfied.  
For example, in the Adult Census dataset ($n = 32{,}560$), this constraint allows a maximum fingerprint length of $L = 2035$ (via choosing maximum $1/\gamma = 1$).
If $L$ needs to be reduced for better robustness against single-user attacks, it can be upper-bounded by \Cref{eq:tardos-length} using the expected number of recipients $N$, number of expected colluding partners $c$ and a small error value $\epsilon<<0.01$.

Increasing $1/\gamma$ has a negative effect on fidelity and utility, however insignificant. In our experiments, the highest observed fidelity degradation was Hellinger distance of $0.0225$ and KL divergence of $6\times10^{-3}$, and the maximum decrease in utility was a drop of 1.6\% in classification accuracy, indicating that even high embedding ratios are acceptable in practice.
Hence, the recommendation is to choose $1/\gamma$ from the higher range, close to the maximum, 1.0, leaving an opportunity to maximise $L$. 
On the other hand, reducing $1/\gamma$ improves fidelity and utility but lowers robustness at the rate shown in \Cref{fig:robustness-subset,fig:robustness-flipping} along the color-axis. 

The neighbourhood size \textit{k} controls the local context for density estimation and correlation preservation. 
We observe that smaller \textit{k} improves effectiveness and fidelity by focusing on highly similar records, while larger \textit{k} improves robustness by stabilising density estimates against data shifts. 
The negative impact of smaller \textit{k} on robustness is minor and can typically be offset by tuning $\omega$. 
Based on experiments, we recommend setting $k\leq1\%$ of the dataset size. 
 
\section{Conclusion}\label{sec:conclusions}
In this work, we introduced NCorr-FP, a novel method for fingerprinting structured data that enables ownership verification and tracing data copies. 
Our method preserves the correlations and value combinations present in the original data. 
Empirical results show that fingerprints are virtually imperceptible, with Hellinger distances below 0.023 and KL divergences below $6\times 10^{-3}$, even at high embedding ratios.
Hence, the fingerprints exhibit better imperceptibility compared to the prior works. 
The method demonstrates strong robustness against a wide range of data-modifying attacks, including those launched by informed adversaries targeting influential records.

We further discussed the integration of a critical component into our fingerprinting system, the probabilistic collusion resolution.  
Evaluated on a mixed-type, open-source dataset, our system maintains high effectiveness and utility.  
The precision of collusion resolution reaches $\geq0.88$ when up to 10 recipients are colluding in various data-merging strategies. 
This is achieved for a fingerprint length that, at the same time, shows high resilience against single-user attacks. 
Collusion success can be improved even further with longer fingerprints, however at the cost of reducing robustness.
To support the practical applicability of the system, we provided a summary of the main fingerprinting parameters and their trade-offs with respect to effectiveness, fidelity, utility, and robustness. 
In future work, our goal is to develop guidelines tailored to specific dataset properties, such as data type distributions, correlation structures, and application contexts. 

\bibliographystyle{ieeetr}
\bibliography{references}  
\end{document}